\documentclass[twocolumn]{aastex631}

\usepackage{amsmath}
\usepackage{relsize}

\received{June 8, 2022}
\revised{August 31, 2022}
\accepted{September 2, 2022}

\shorttitle{Isolated YSOs}

\graphicspath{{./}{figures/}}

\begin{document}

\title{Spectroscopic Confirmation of a Population of Isolated, Intermediate-Mass YSOs}

\submitjournal{AJ}

\shortauthors{Kuhn et al.}

\author[0000-0002-0631-7514]{Michael A. Kuhn}
\affil{Department of Astronomy, California Institute of Technology, 1216 East California Boulevard, Pasadena, CA 91125, USA}

\author[0000-0003-2015-3429]{Ramzi Saber}
\affil{Department of Astronomy, California Institute of Technology, 1216 East California Boulevard, Pasadena, CA 91125, USA}

\author[0000-0001-9062-3583]{Matthew S. Povich}
\affil{Department of Physics and Astronomy, California State Polytechnic University, 3801 West Temple Ave, Pomona, CA 91768, USA}

\author[0000-0001-7207-4584]{Rafael S. de Souza}
\affiliation{Key Laboratory for Research in Galaxies and Cosmology, Shanghai Astronomical Observatory,\\
Chinese Academy of Sciences, 80 Nandan Rd., Shanghai 200030, China}

\author[0000-0002-2308-6623]{Alberto Krone-Martins}
\affiliation{Donald Bren School of Information and Computer Sciences, University of California, Irvine, CA 92697, USA}
\affiliation{CENTRA/SIM, Faculdade de Ci\^{e}ncias, Universidade de Lisboa, Ed. C8, Campo Grande, 1749-016, Lisboa, Portugal}

\author[0000-0002-0406-076X]{Emille E. O. Ishida}
\affiliation{Universit\'{e} Clermont Auvergne, CNRS/IN2P3, LPC, F-63000 Clermont-Ferrand, France}

\author[0000-0002-2250-730X]{Catherine Zucker}
\affiliation{Space Telescope Science Institute, 3700 San Martin Drive, Baltimore, MD 21218, USA}
\affiliation{NASA Hubble Fellowship Program}

\author[0000-0002-8109-2642]{Robert A. Benjamin}
\affiliation{Department of Physics, University of Wisconsin-Whitewater, 800 W Main St, Whitewater, WI 53190 USA}

\author{Lynne A. Hillenbrand}
\affil{Department of Astronomy, California Institute of Technology, 1216 East California Boulevard, Pasadena, CA 91125, USA}

\author[0000-0002-9419-3725]{Alfred Castro-Ginard}
\affiliation{Leiden Observatory, Leiden University, Niels Bohrweg 2, 2333 CA Leiden, Netherlands}

\author{Xingyu Zhou}
\affiliation{Kavli Institute for Astronomy and Astrophysics, Peking University, Yiheyuan 5, Haidian Qu, 100871 Beijing, China}
\affiliation{Department of Astronomy, Peking University, Yiheyuan 5, Haidian Qu, 100871 Beijing, China}

\correspondingauthor{Michael A. Kuhn}
\email{mkuhn@astro.caltech.edu}

\collaboration{999}{for the COIN collaboration}

\begin{abstract}
Wide-field searches for young stellar objects (YSOs) can place useful constraints on the prevalence of clustered versus distributed star formation. The Spitzer/IRAC Candidate YSO (SPICY) catalog is one of the largest compilations of such objects ($\sim$120,000 candidates in the Galactic midplane). Many SPICY candidates are spatially clustered, but, perhaps surprisingly, approximately half the candidates appear spatially distributed. To better characterize this unexpected population and confirm its nature, we obtained Palomar/DBSP spectroscopy for 26 of the optically-bright ($G<15$~mag) ``isolated'' YSO candidates. We confirm the YSO classifications of all 26 sources based on their positions on the Hertzsprung-Russell diagram, H and Ca\,{\sc ii} line-emission from over half the sample, and robust detection of infrared excesses. This implies a contamination rate of $<$10\% for SPICY stars that meet our optical selection criteria. Spectral types range from B4 to K3, with A-type stars most common. Spectral energy distributions, diffuse interstellar bands, and Galactic extinction maps indicate moderate to high extinction. Stellar masses range from $\sim$1--7~$M_\odot$, and the estimated accretion rates, ranging from $3\times10^{-8}$ to $3\times10^{-7}$~$M_\odot$\,yr$^{-1}$, are typical for YSOs in this mass range. The 3D spatial distribution of these stars, based on Gaia astrometry, reveals that the ``isolated'' YSOs are not evenly distributed in the Solar neighborhood but are concentrated in kpc-scale dusty Galactic structures that also contain the majority of the SPICY YSO clusters. Thus, the processes that produce large Galactic star-forming structures may yield nearly as many distributed as clustered YSOs.  
\end{abstract}


\defcitealias{SPICY}{Paper~I}
\defcitealias{Kuhn2021_sgr}{Paper~II}

\section{Introduction} \label{sec:intro}

Young stellar objects (YSOs) exhibit a high degree of spatial clustering \citep{2000AJ....120.3139C,2003ARA&A..41...57L,2014ApJ...787..107K,2022PASP..134d2001M}, which arises as a consequence of the hierarchical structure of the molecular clouds from which they form
\citep{2007ARA&A..45..565M,2019ARA&A..57..227K,2021MNRAS.506.3239G}. 
Most known YSOs are found in star-forming complexes that may contain hundreds to thousands of stars \citep{2008hsf1.book.....R,2008hsf2.book.....R}. Theoretical models of star formation on a Galactic scale tend to focus on massive star-forming regions, which are thought to domination the star-formation rate \citep[e.g.,][]{2014prpl.conf..291L,2020MNRAS.494..624K}.

Nevertheless, there are many examples of YSOs that appear to be in relative isolation or small groups. These range from the nearest YSOs in the TW Hya Association \citep{2004ARA&A..42..685Z}, to small clusters around Herbig Ae/Be stars \citep{1995AJ....109..280H}, molecular clouds with low stellar yields \citep{2008hsf1.book...18P,2009ApJ...704..292F}, and O stars that appear to be located outside major star-forming regions \citep{2004AJ....127.1632O}. 

The prevalence of clustered versus distributed star formation has implications for star-formation theory. For example, supersonic turbulence in molecular clouds, proposed as a possible regulatory mechanism for the star-formation rate of the Galaxy \citep[][]{2005ApJ...630..250K}, is expected to influence the spatial distribution of young stars. Strong turbulent pressure support would produce isolated star-forming cores, while dense clusters could form more easily in its absence \citep{2004RvMP...76..125M}. 

Astronomical surveys that provide coverage over large areas of the sky can be useful when searching for distributed YSOs outside known star-forming regions. However, in these searches, contamination rates are expected to be higher for isolated YSO candidates than for clustered YSO candidates, given that contaminants tend to be uniformly distributed \citep[e.g.,][]{2018A&A...615L...1M,2020ApJ...899..128K}. 

The Spitzer/IRAC Candidate YSO (SPICY) catalog \citep[][hereafter Paper~I]{SPICY} is one of the most extensive censuses of YSOs in the Galactic midplane, the region of the Galaxy with the highest star-formation activity. This catalog contains $\sim$10$^5$ candidate YSOs, selected by their mid-infrared properties from the Galactic Legacy Infrared Mid-Plane Survey Extraordinaire \citep[GLIMPSE;][]{2003PASP..115..953B,2009PASP..121..213C}, and related projects. 

As expected, the SPICY YSOs trace out numerous young star clusters, including both previously known and newly identified star-forming regions \citep[][hereafter Paper~II]{Kuhn2021_sgr}. More surprisingly, nearly 50\% of the candidate YSOs do not appear spatially associated with other young stars. If this ratio of clustered-to-distributed YSOs holds, it would imply that distributed star formation contributes significantly to the Galaxy's star-formation rate. Other survey-based censuses have also reported significant numbers of non-clustered young stellar candidates
\citep[e.g.,][]{2008AJ....136.2413R,2016MNRAS.458.3479M,2019MNRAS.487.2522M,2018A&A...620A.172Z,Vioque2020,2020AJ....160...68W,2021AJ....162..282M}. In these studies, candidate identification is typically based on photometric data and, more recently, on astrometric data from the Gaia mission \citep{2016A&A...595A...1G}. Thus, follow-up observations are necessary for confirmation of YSOs. Here, we spectroscopically examine a sample of 26 YSO candidates from SPICY to test whether there is evidence for an isolated population in the Galaxy.

The remainder of the paper is organized as follows. Section~\ref{sec:spicy} summarizes the SPICY catalog and describes our criteria for selecting program stars. Section~\ref{sec:observations} describes observations and data reduction. Section~\ref{sec:properties} provides inferred stellar properties, including spectral types, emission lines, extinctions, luminosities, masses, ages, and accretion rates. Section~\ref{sec:validation} presents our arguments that the program stars are all YSOs and uses this sample to constrain SPICY contamination rates. Section~\ref{sec:isolation} examines the Galactic environments of the program stars. Finally, Section~\ref{sec:dc} provides our discussion and conclusions. 

\begin{figure*}[t]
\centering
\includegraphics[angle=0.,width=0.3\textwidth]{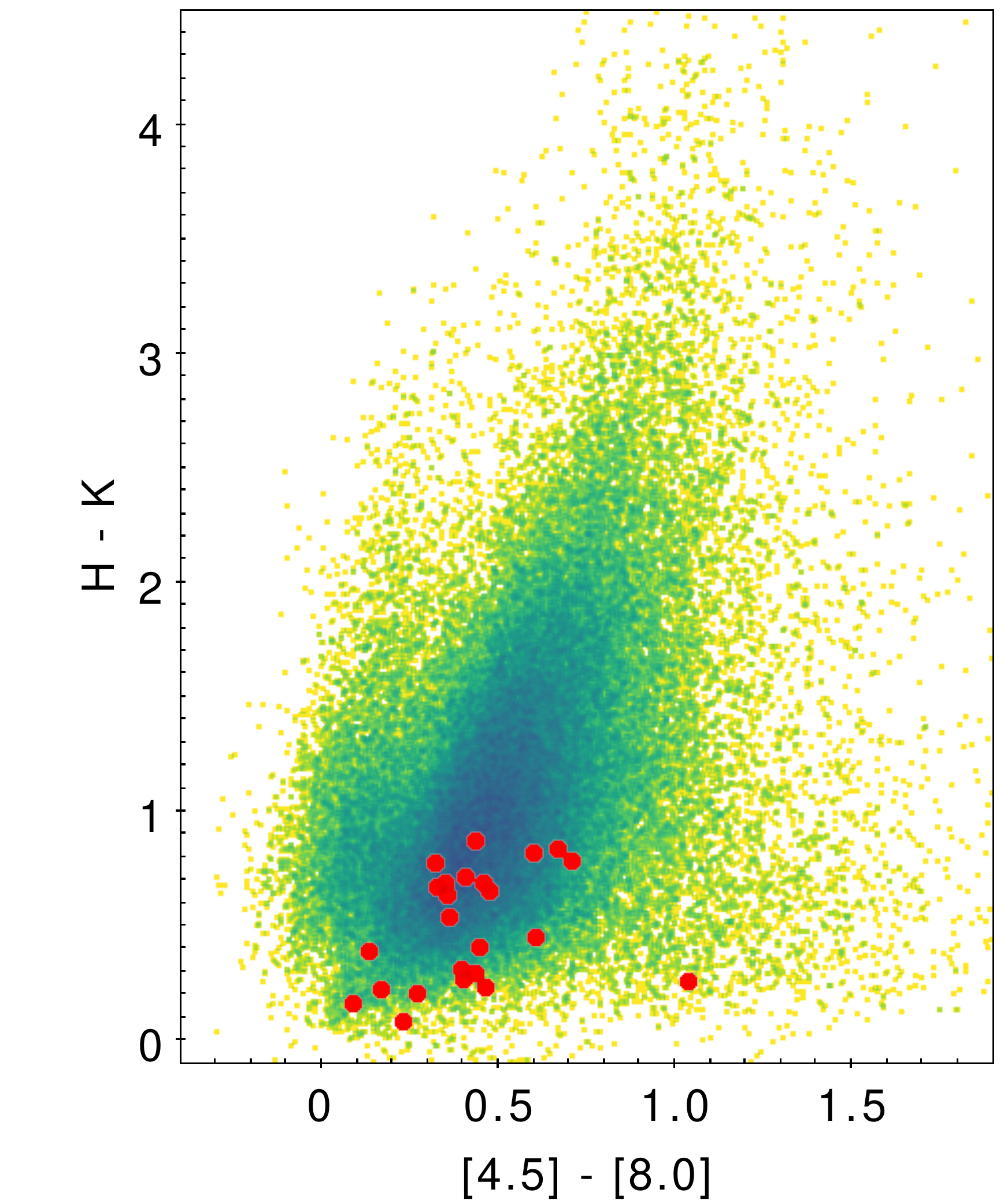}
\includegraphics[angle=0.,width=0.3\textwidth]{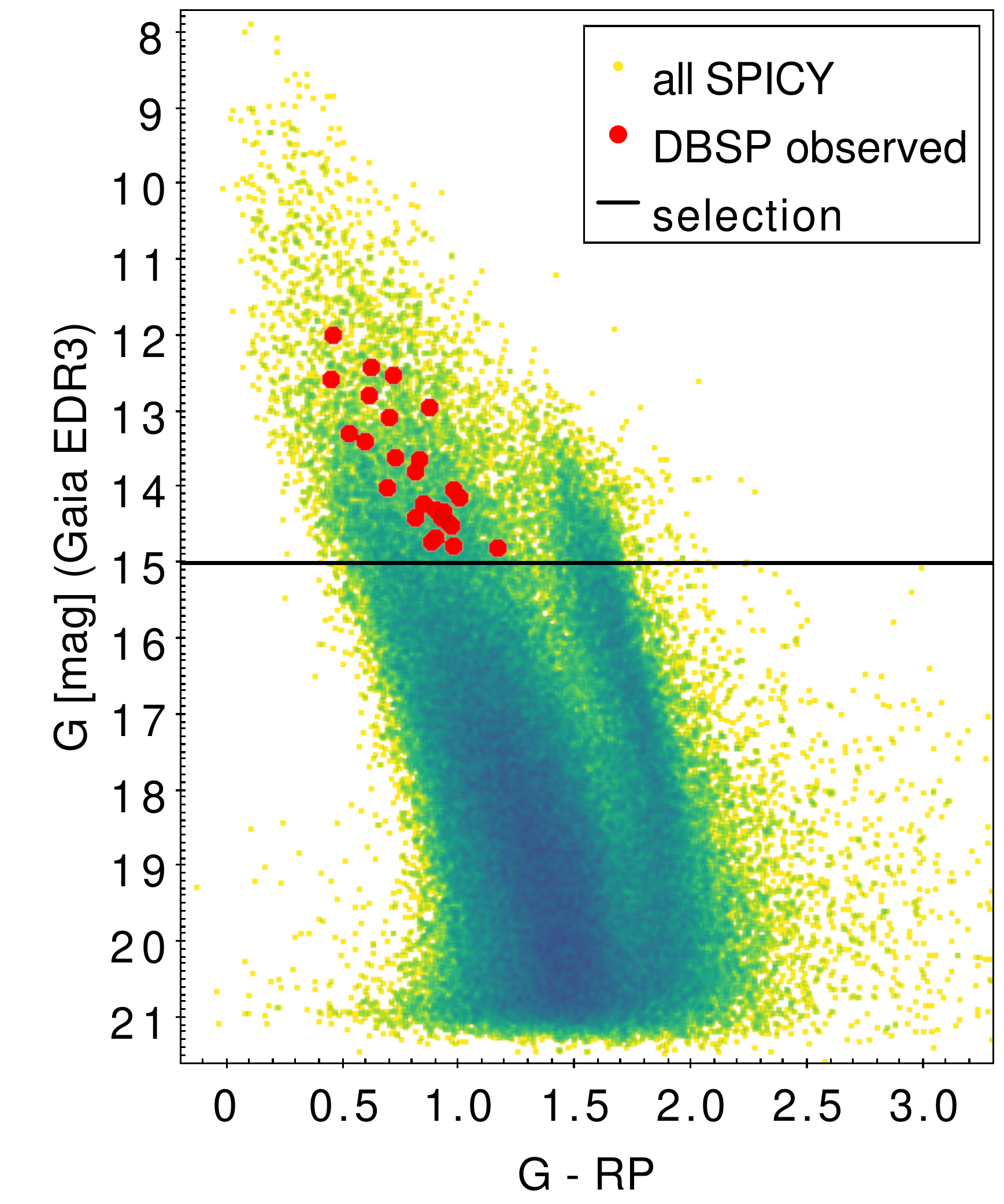}
\includegraphics[angle=0.,width=0.3\textwidth]{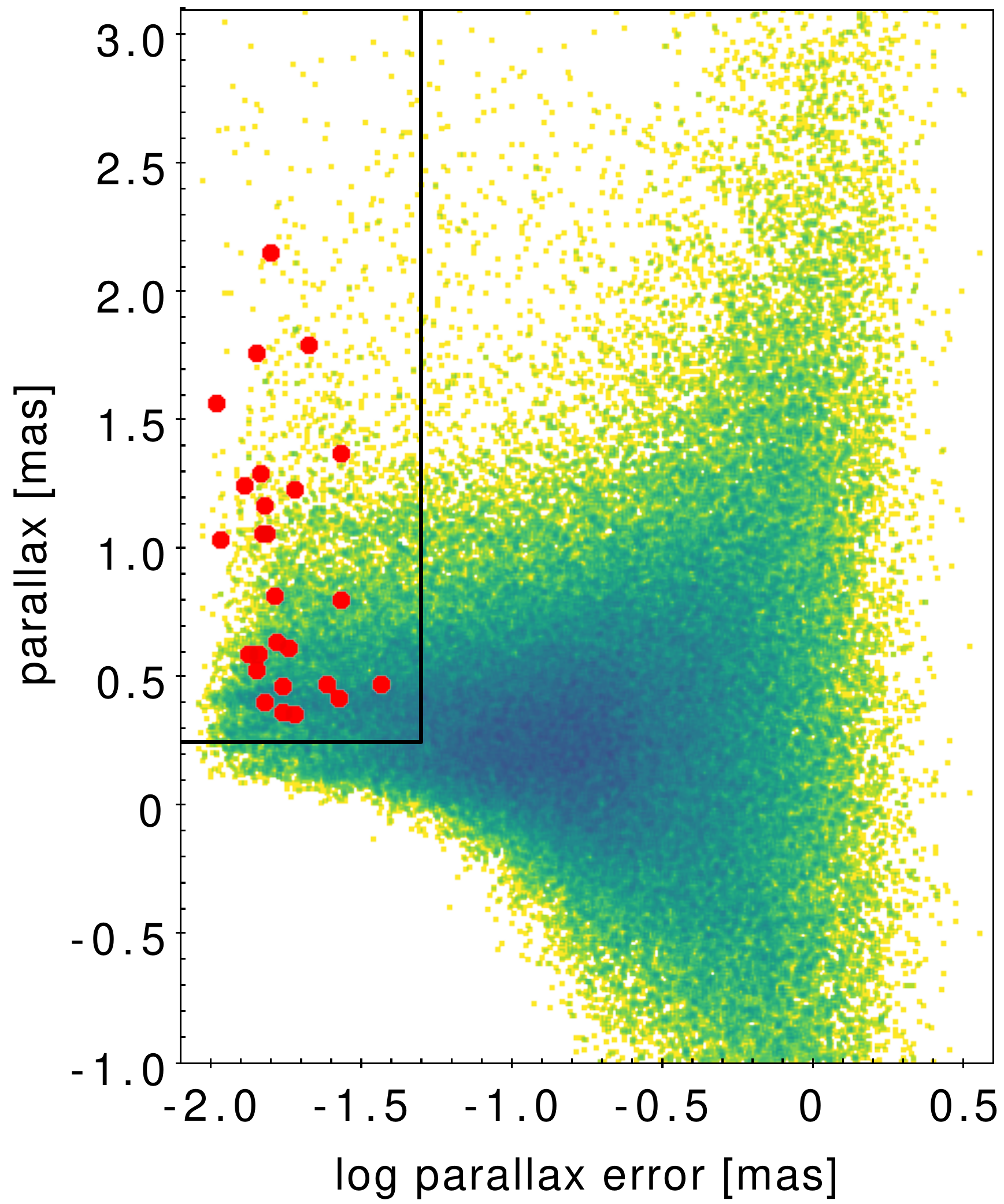}
\caption{Comparison of the photometric and astrometric properties of the program stars (red points) observed with Palomar/DBSP to the full SPICY catalog (green points). Left: Near-infrared $H-K$ versus Spitzer/IRAC $[4.5]-[8.0]$ color-color diagram. Center: Gaia $G$-band magnitude versus $G - G_{RP}$. Right: Gaia parallax versus parallax error (log scale). The black lines indicate the cuts used for selecting stars for follow-up.  
\label{fig:sample}}
\end{figure*}

\section{The SPICY Catalog}\label{sec:spicy}

The SPICY catalog \citepalias{SPICY} contains 117,446 candidate YSOs identified using mid-infrared photometry from Spitzer/IRAC observations of the Galactic midplane, including the Galactic Legacy Infrared Mid-Plane Survey Extraordinaire (GLIMPSE) I, II, 3D, Galactic Center, Vela-Carina, Cygnus X, and Spitzer Mapping of the Outer Galaxy surveys \citep{2003PASP..115..953B,2006JPhCS..54..176S,2009PASP..121..213C,2007sptz.prop40791M,2009ApJ...707..510Z,2010ApJ...720..679B,2019ApJ...880....9W,2020AJ....160...68W}. These were analyzed in tandem with the near-infrared photometry from the Two-Micron All Sky Survey \citep[2MASS;][]{2006AJ....131.1163S}, the United Kingdom Infrared Telescope (UKIRT) Infrared Deep Sky Survey \citep[UKIDSS;][]{2007MNRAS.379.1599L,2008MNRAS.391..136L}, and the Visible and Infrared Survey Telescope for Astronomy (VISTA) Variables in the V\'ia L\'actea \citep[VVV;][]{2010NewA...15..433M,2018MNRAS.474.1826S}. Altogether, the SPICY YSO catalog covers 613 square degrees near the Galactic midplane between Galactic coordinates $-105^\circ\leq\ell\leq 110^\circ$ and $|b|\lesssim1$--2$^\circ$.

Candidate YSOs were identified based on excess infrared emission consistent with the spectral energy distributions (SEDs) of pre-main-sequence stars with disks or envelopes, using the YSOs analyzed by \citet{Povich2013} as templates. A random-forest classifier was employed \citep{ho1995random,breiman2001random}, and only the near- and mid-infrared photometry, without consideration of sky coordinates, were used as features. Thus, the catalog provides a view of the distribution of YSOs in the Galactic midplane that is nearly spatially unbiased, apart from issues of Spitzer survey sensitivity in regions with differing levels of point-source crowding and mid-infrared nebulosity \citep{2013ApJS..207....9K}. Infrared observations of star-forming regions tend to unveil deeply embedded YSOs that are not optically visible. In the case of SPICY, only about one third of the stars are detected by Gaia.

\subsection{Clustered and Distributed Stars in SPICY}\label{sec:definition}

The candidate YSOs in the SPICY catalog include both clustered and distributed stars. \citetalias{SPICY} performed cluster analysis on the Galactic $(\ell,b)$ coordinates of this sample, finding over 400 groups containing at least 30 stars via the \texttt{HDBSCAN} algorithm \citep{campello2013density} using the ``Excess of Mass'' method.\footnote{This initial cluster analysis was based only on angular positions in, since Gaia astrometry is unavailable for two thirds of the SPICY sample.} 
These groups include both stellar clusters and stellar associations, and their members collectively account for half the YSO candidates \citepalias[][their Section~7]{SPICY}. 
In \citetalias{Kuhn2021_sgr}, Gaia reveals that optically visible members of the same SPICY groups tend to have similar parallaxes and proper motions, providing corroboration that they are physically associated young stars. However, this strategy cannot be used to verify the spatially distributed stars in the SPICY catalog, so other methods must be used to confirm their youth.
   
Here, we base our samples of ``clustered'' and ``isolated'' SPICY stars on whether the sources are included in the groups from \citetalias{SPICY}. This division is algorithm dependent, since different cluster analysis algorithms may yield different results \citep{Everett2011}. Nevertheless, the HDBSCAN algorithm has become a popular method for identifying young stellar clusters and associations \citep[e.g.,][]{2019A&A...626A..80C,Kounkel2020,Kerr2021}. 
It is also possible that some ``isolated'' SPICY stars may be members of stellar groups not detected in Paper~I, which could include clusters with low disk fractions or whose members are mostly too faint to be detected. These possibilities are investigated in Section~\ref{sec:isolation}.
 
\begin{figure*}[t]
\centering
\includegraphics[angle=0.,width=1\textwidth]{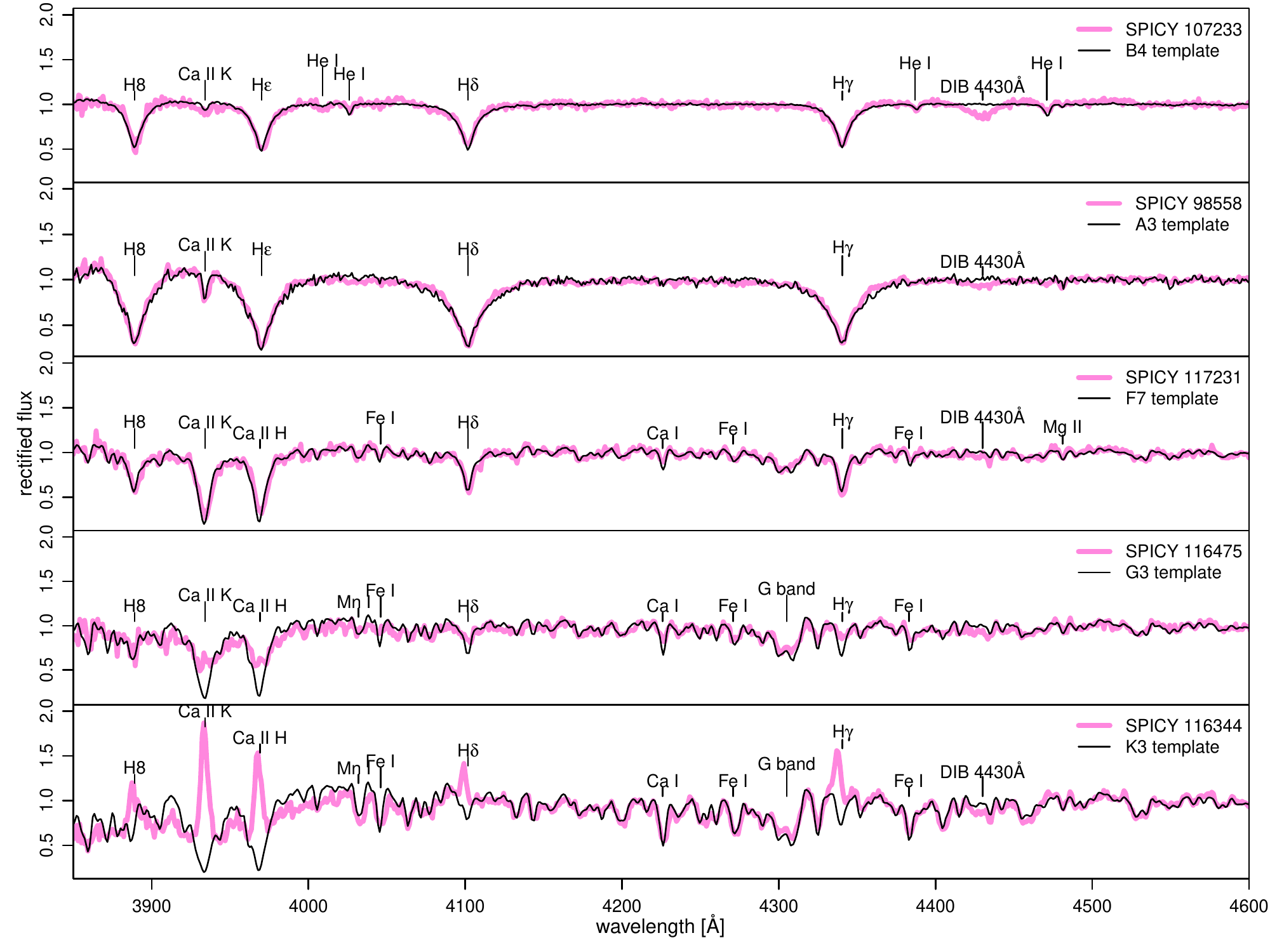}
\caption{Spectra of several program stars from 3850--4600~\AA. The pink line is the observed spectrum, and the best-fitting template (A6) is overplotted in black. Several notable spectral features are marked.\\
(The complete figure set (6 panels) is available.)
\label{fig:spectra}}
\end{figure*} 
   
\subsection{Spectroscopic Follow-up Sample}\label{sec:sample}   

We selected stars for follow-up that are both sufficiently bright at optical wavelengths for spectroscopic analysis and have high-quality astrometric data from Gaia's early third data release \citep[EDR3;][]{2021A&A...649A...1G}. The criteria were the following: Gaia $G < 15$~mag, parallax $\varpi>0.25$~mas, parallax uncertainty $\sigma_\varpi < 0.05$~mas, renormalized unit weight error $\mathrm{RUWE} <1.4$, and YSO Class~II or earlier in the SPICY catalog. There are 806 stars in the entire SPICY catalog that meet these criteria.

The approximate parity between clustered to non-clustered stars in the full SPICY sample extends to the sources that meet these selection criteria. Of the 806 stars, 351 are members of \texttt{HDBSCAN} groups while 455 are not -- our program stars are selected from the latter set. We obtained spectroscopy for 26 of these objects (Table~\ref{tab:log}), which comprised the majority of those visible from Palomar Observatory at the time of observation.

Some properties of the program stars, relative to the full catalog, are shown in Figure~\ref{fig:sample}. These stars represent the bright end of the distribution, and, on average, they have bluer Gaia $G-G_{RP}$ colors and have smaller distances. The SPICY stars with the reddest $H-K$ colors are not represented. However, in the mid-infrared, the sample spans nearly the full range of $[4.5]-[8.0]$ colors found in SPICY.

Although our program stars are relatively bright, most have received little previous attention. Several have appeared in lists of variable or emission-line stars or as ``intrinsically red'' sources (Appendix~\ref{appendix:individual}), but the nature of most of these objects has remained uncertain. 

In contrast, three program stars on our list were included in a study by \citet{Vioque2022} (published while this paper was in preparation), in which they spectroscopically classified 128 Herbig Ae/Be stars, more than doubling the known members of this class. Their lists were based on candidates derived from optical (including H$\alpha$) and infrared photometry \citep{Vioque2020}. The three overlapping stars were all classified as Herbig~Ae/Be stars. Comparison of our spectral types with those from \citet{Vioque2022} is in Appendix~\ref{appendix:individual}. 

\section{Spectroscopic Observations and Data Reduction}\label{sec:observations}

The stars were observed using the Double Spectrograph \citep[DBSP;][]{1982PASP...94..586O} on the 200-inch Hale Telescope at Palomar Observatory on June 5th, 2021 UT (Table~\ref{tab:log}). The observations were taken using the D68 dichroic, with the 600~line/mm grating blazed at 3780~\AA\ for the blue side and the 1200~line/mm grating blazed at 7100~\AA\ for the red side. Observations were made with the $1^{\prime\prime}$ slit. This gave a wavelength range of 3816--6874~\AA\ and 7381--9015~\AA\ for the two spectrographs, with resolutions of $R\approx 2000$ and $R\approx 6000$, respectively. On the red side, the wavelength range 7590-7710~\AA\ was unusable due to detector noise. 
Exposure times were selected to achieve signal-to-noise ratios of $\sim$50 on the red side, yielding signal-to-noise ratios a factor of a few higher on the blue side. Data were reduced using the \texttt{DBSP\_DRP} pipeline \citep{2021arXiv210712339R,2021ascl.soft08020R}. The standard stars BD+28 4211 and Feige~34 were used for spectral flux calibration. 

\section{Inference of Stellar Properties}\label{sec:properties}

\subsection{Spectral Classification}\label{sec:spectra}

Spectral classes were inferred through comparison of the DBSP spectra to templates from a library of co-added standard-star spectra observed by the BOSS spectrograph \citep{Kesseli2017}. The library has a resolution of $R\approx2000$, providing a good match for the blue-side program star spectra. 

\begin{figure}[t]
\centering
\includegraphics[angle=0.,width=0.48\textwidth]{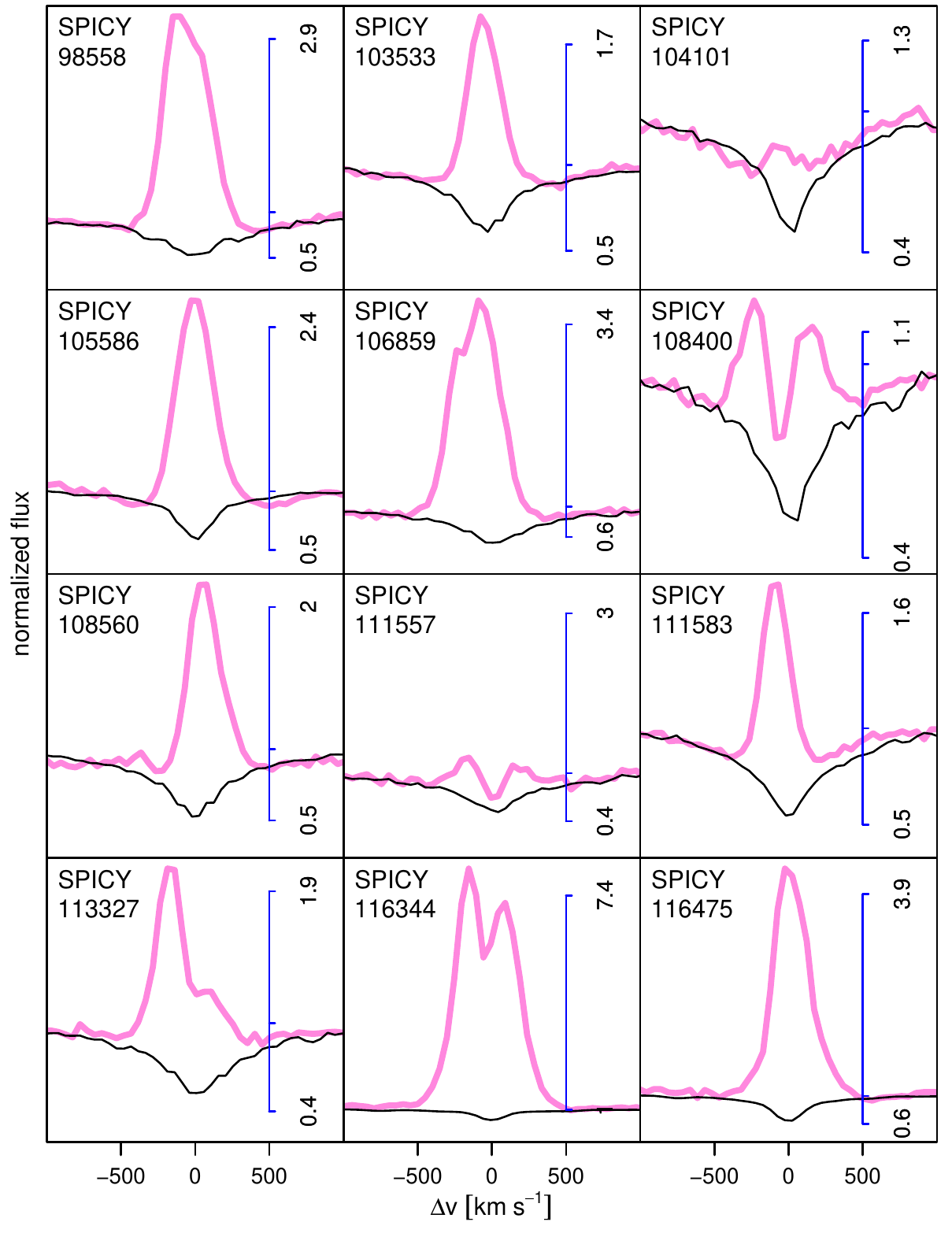}
\caption{Line profiles for program stars with clear H$\alpha$ emission. The observed spectra are shown in magenta, and the corresponding templates are shown in black. Spectra are plotted in continuum units, as indicated by the blue scale bar. 
\label{fig:Ha}}
\end{figure}

Spectral types were assigned using $\chi^2$ fitting of the blue-side spectra (Figure~\ref{fig:spectra}). First, both template and program-star spectra were normalized by fitting their continua with smooth non-parametric curves, then dividing the spectra by these fits. For continuum fitting we used the local regression \citep[{\tt loess};][]{Cleveland1979}, with the $\texttt{span}$ parameter set to 0.15. The $\chi^2$ statistic was then computed for each template/program-star pair, and the template producing the minimum value was adopted. We omitted the spectral region around H$\alpha$ (Figure~\ref{fig:Ha}), along with any other regions with evidence of line emission. The formal uncertainties derived from the $\chi^2$ statistic underestimated the uncertainty in temperature class, so we visually inspected the model fits to determine the templates that bracket each star's spectral type, focusing on diagnostic features suggested by \citet{2009ssc..book.....G}. 

Inferred temperature classes are provided in Table~\ref{tab:spec}. Spectral types range from B4 to K3; however, the most common spectral types are A (12 examples) and F (8 examples). Spectral types are converted to effective temperatures, $T_\mathrm{eff}$, using the temperature scale for young stars derived by \citet[][]{PecautMamajek2013}.\footnote{
We use temperatures from Table~6 in \citet[][]{PecautMamajek2013} for F0 or later and Table~5 for A9 and earlier.} The \citet{Kesseli2017} library subdivides the template spectra into ``dwarfs'' ($6\geq \log g \geq 3.8$) and ``giants'' ($2\leq \log g \leq 3.2$). On their own, our low-resolution spectra do not provide strong constraints on gravity. All program star spectra are consistent with the ``dwarf'' templates, except for SPICY~104101 where a ``giant'' template provided a better match.\footnote{Given the mass and radius assigned to SPICY~104101 in Section~\ref{sec:mass}, its $\log g = 3.2$ is consistent with the ``giant'' category.} We assume Solar metallicity for all objects except SPICY~90918, where $\mathrm{[Fe/H]} = +0.5$ provides a substantially better fit.

\begin{figure}[t]
\centering
\includegraphics[angle=0.,width=0.48\textwidth]{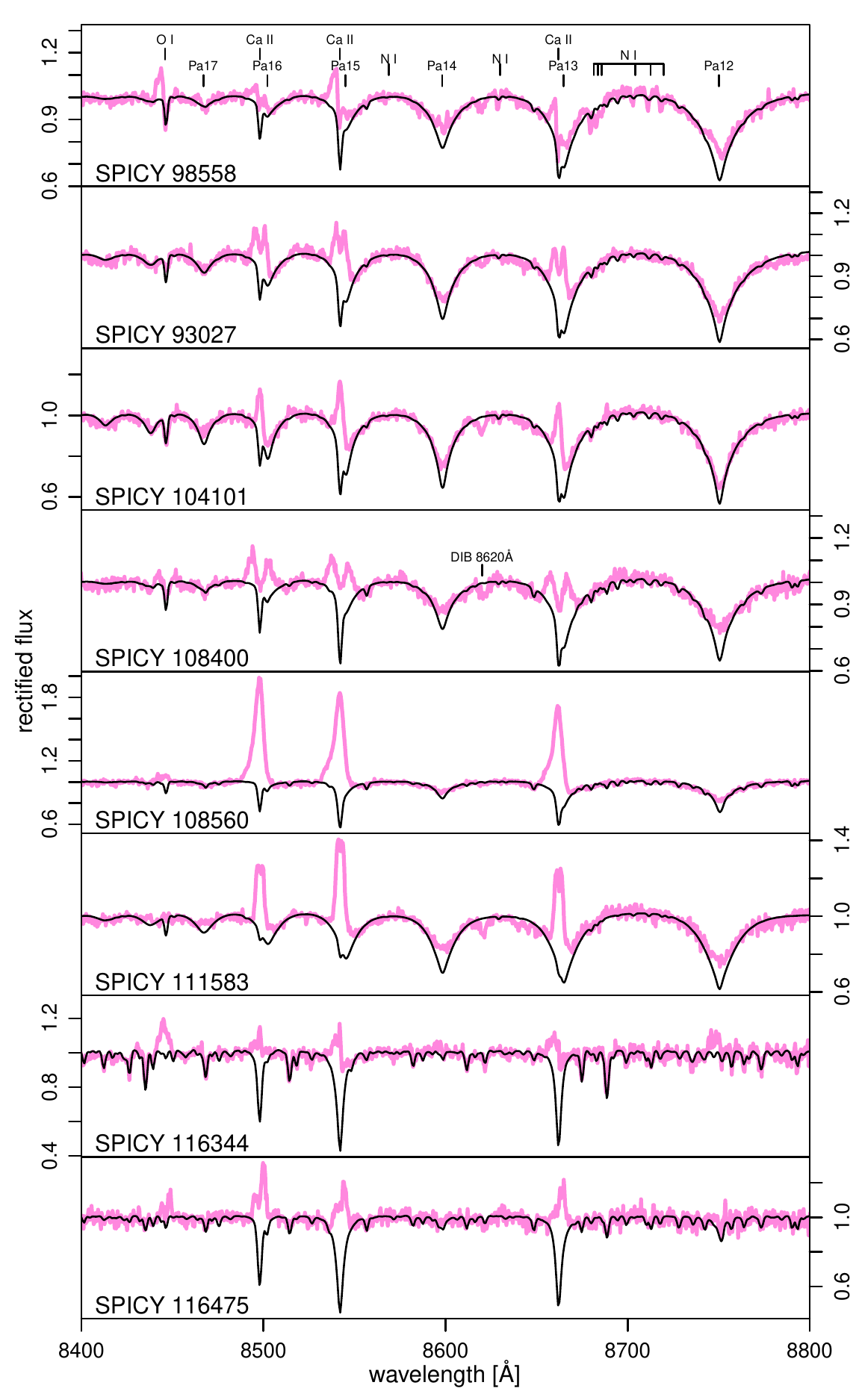}
\caption{Spectra of select program stars (magenta curves) with emission in the Ca\,{\sc ii} 8498~\AA, 8542~\AA, and 8662~\AA\ lines. Other lines occasionally seen in emission include hydrogen lines (Pa~12-17) and the O\,{\sc i} line at 8446~\AA. The BOAZ model with the closest match to our adopted stellar parameters is shown in black.\\
(The complete figure set (5 panels) is available.)
\label{fig:CaII}}
\end{figure}

The red-side spectra are shown in Figure~\ref{fig:CaII}, focusing on the region around the infrared Ca\,{\sc ii} triplet. 
The \citet{Kesseli2017} templates are noisy in this region, so we compare the program-star spectra to the BOAZ models \citep{2012AJ....144..120M,Bohlin2017}. For each program star, we plotted the model from the BOAZ grid ($R=5000$ resolution) that most closely agrees with the program star's classification, including $T_\mathrm{eff}$ (derived from the blue side) and $\log g$ (derived in Section~\ref{sec:sed}, below). Overall, the selected models are consistent with the features seen in these spectra, including the Pa12-17 and Ca\,{\sc ii} triplet lines (when not in emission). Remarks about spectral features of individual stars are given in Appendix~\ref{appendix:individual}.

Veiling may provide another systematic source of uncertainty. For cooler stars, spectral types derived without accounting for veiling may be up to 4 subtypes too early \citep{Fang2020}. However, this effect is not expected to be as significant for the hotter stars in this sample, where the stellar photosphere is more dominant.  

\vspace*{-0.10cm}
\subsection{Emission Lines}

Several SPICY stars have prominent hydrogen emission lines (Figure~\ref{fig:Ha}), and most other stars have some emission component partially filling in the photospheric absorption lines. Emission lines are also detected from the Ca\,{\sc ii} triplet (Figure~\ref{fig:CaII}), and occasionally from the Ca\,{\sc ii} H and K lines and the O\,{\sc i} line at 8446~\AA. All these emission lines are indicators of accretion for young stars \citep{Hamann1992,Muzerolle1998,2017MNRAS.464.4721F}. Equivalent widths are measured by subtracting the normalized spectrum of the program stars from the normalized best-fitting template spectra, then integrating over the lines (Table~\ref{tab:ew}). 

The prominent H$\alpha$ emission lines have full-widths at 10\% of peak height ($FW10\%$) ranging from 440--830~km~s$^{-1}$. For comparison, our spectral resolution is 150~km~s$^{-1}$, so these lines are resolved. \citet{WhiteBasri2003} estimated a cutoff of $FW10\% > 270$~km~s$^{-1}$ ($\approx$6~\AA) for accreting T Tauri stars, and all measured line-widths fall above this threshold. 

In our sample, most H$\alpha$ lines are single-peaked, but several, including SPICY~108400, SPICY~111557, SPICY~113327, and SPICY~116344, have double peaks. Similar line morphologies are found in atlases of H$\alpha$ line profiles compiled for T Tauri stars \citep{Reipurth1996} and Herbig Ae/Be stars \citep{Carmona2010}. However, in previous studies \citep[e.g.,][]{1984A&AS...57..285F,Vioque2018,Vioque2022}, approximately 50\% of Herbig stars had double-peaked line profiles, which is somewhat more common than in our sample. In the case of T Tauri stars, theoretical models of H$\alpha$ emission from in-falling accretion streams have been largely successful at reproducing the observed profiles \citep{Wilson2022}.  

It is not surprising that some stars (10 out of 26) have no detected H$\alpha$ emission. Investigations have found that young stars without signatures of ongoing accretion may still have excess mid-infrared emission \citep[e.g.,][]{2005A&A...443..541G,2009ApJ...703..399L,2022ApJ...928..134S}.
The intermediate masses of the stars in our sample mean that they have higher continuum levels than low-mass stars, which can contribute to masking weak emission lines.

\begin{figure*}[t]
\centering
\includegraphics[angle=0.,width=0.9\textwidth]{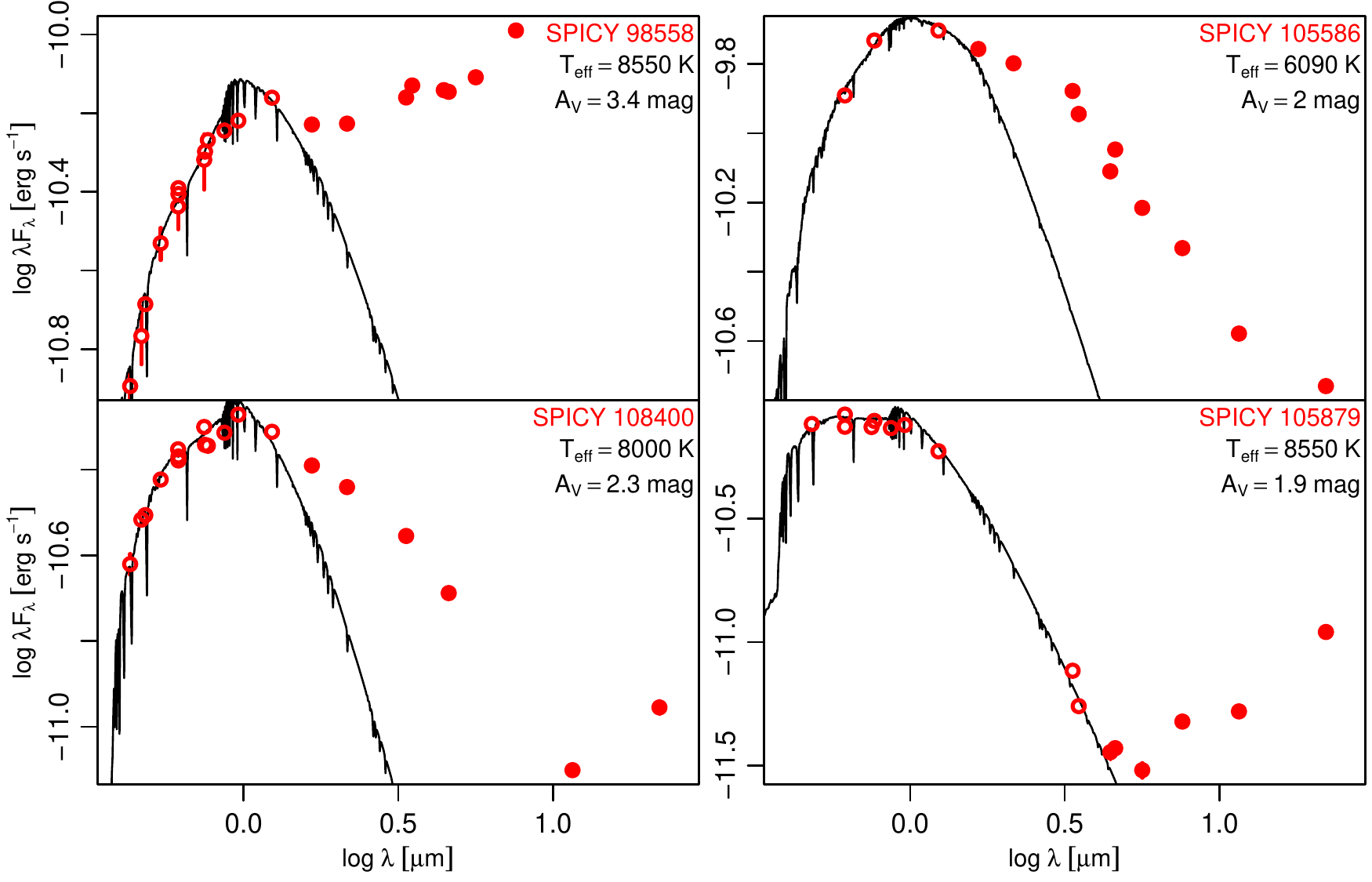}
\caption{SED plots of select SPICY stars. Flux measurements from photometric measurements are plotted as the red circles with error bars. (Error bars smaller than the symbol sizes are not drawn.) Points included in the fit are open circles, whereas points showing infrared excess are filled. The black curve is the reddened stellar photosphere model that provides the best fit to the photometry.\\
(The complete figure set (26 panels) is available.)
\label{fig:sed}}
\end{figure*}

\begin{figure*}[t]
\centering
\includegraphics[angle=0.,width=0.46\textwidth]{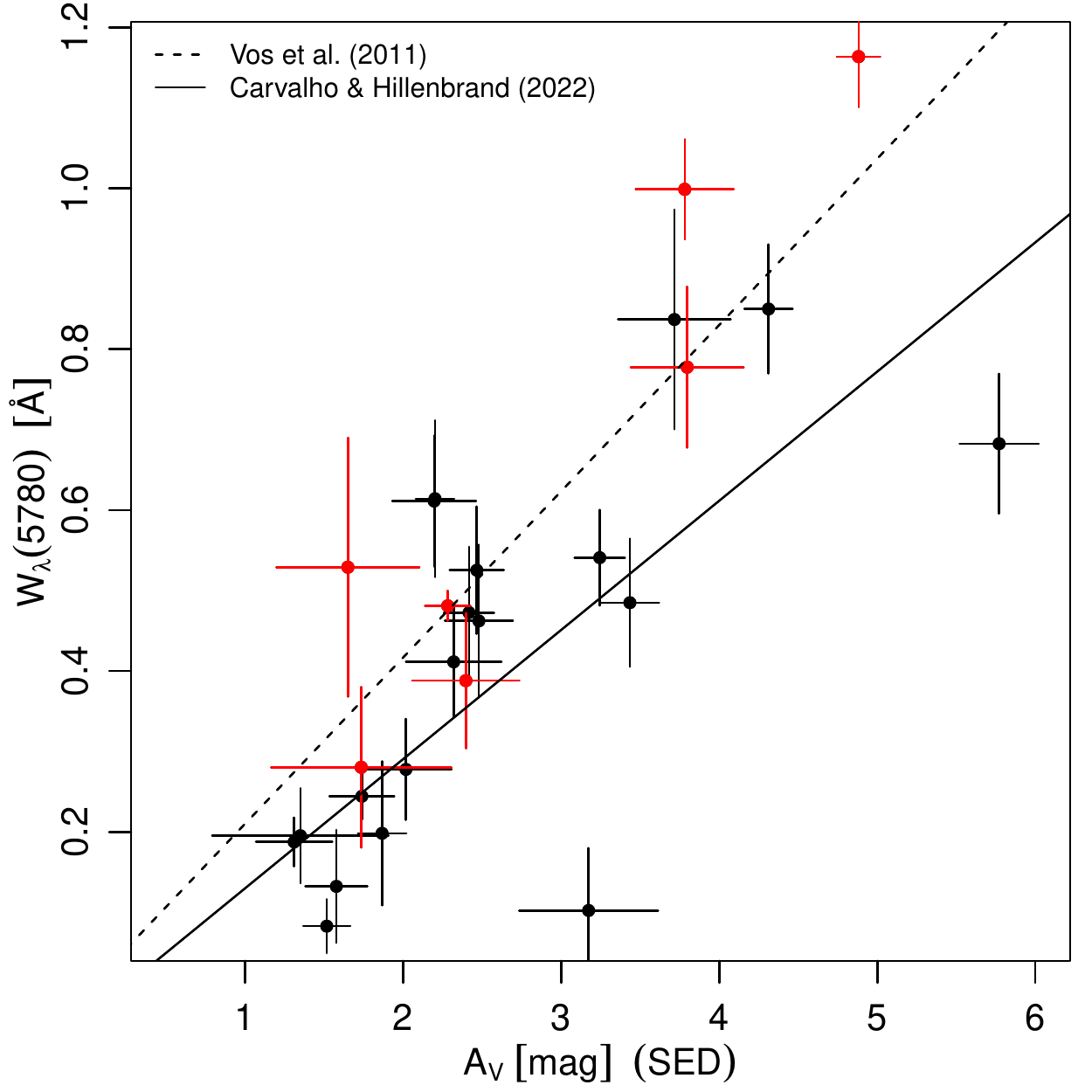}
\includegraphics[angle=0.,width=0.46\textwidth]{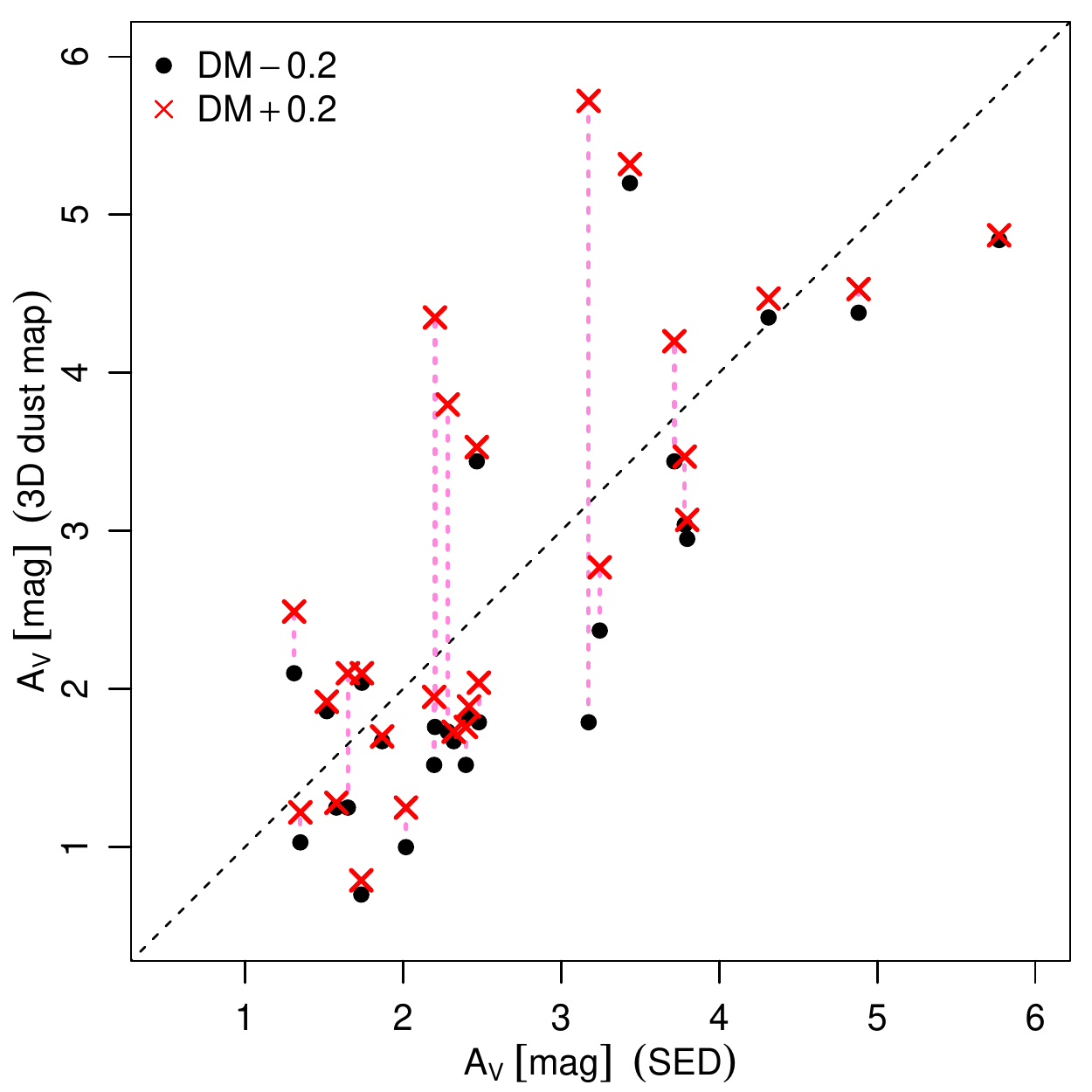}
\caption{Left: The equivalent width of the 5780~\AA\ DIB versus $A_V$ values from SED fitting. The lines indicate relations from \citet{Vos2011}, for ``$\sigma$-type'' sightlines, and from \citet{Carvalho2022}, for ``$\zeta$-type'' sightlines. Sources with $W_\lambda(5780)/W_\lambda(5797) > 3.3$ (indicative of $\sigma$-type sightlines) are red.
Right: Extinctions along the line of sight to each program star from the \citet{2019ApJ...887...93G} Galactic dust map. The ordinate shows extinctions at the stars' distance moduli ($DM$) $-0.2$ and $+0.2$, and the abscissa is the extinction of the star itself. The black, dashed line marks equal $A_V$ values. 
\label{fig:dib1}}
\end{figure*}

\begin{figure}[t]
\centering
\includegraphics[angle=0.,width=0.48\textwidth]{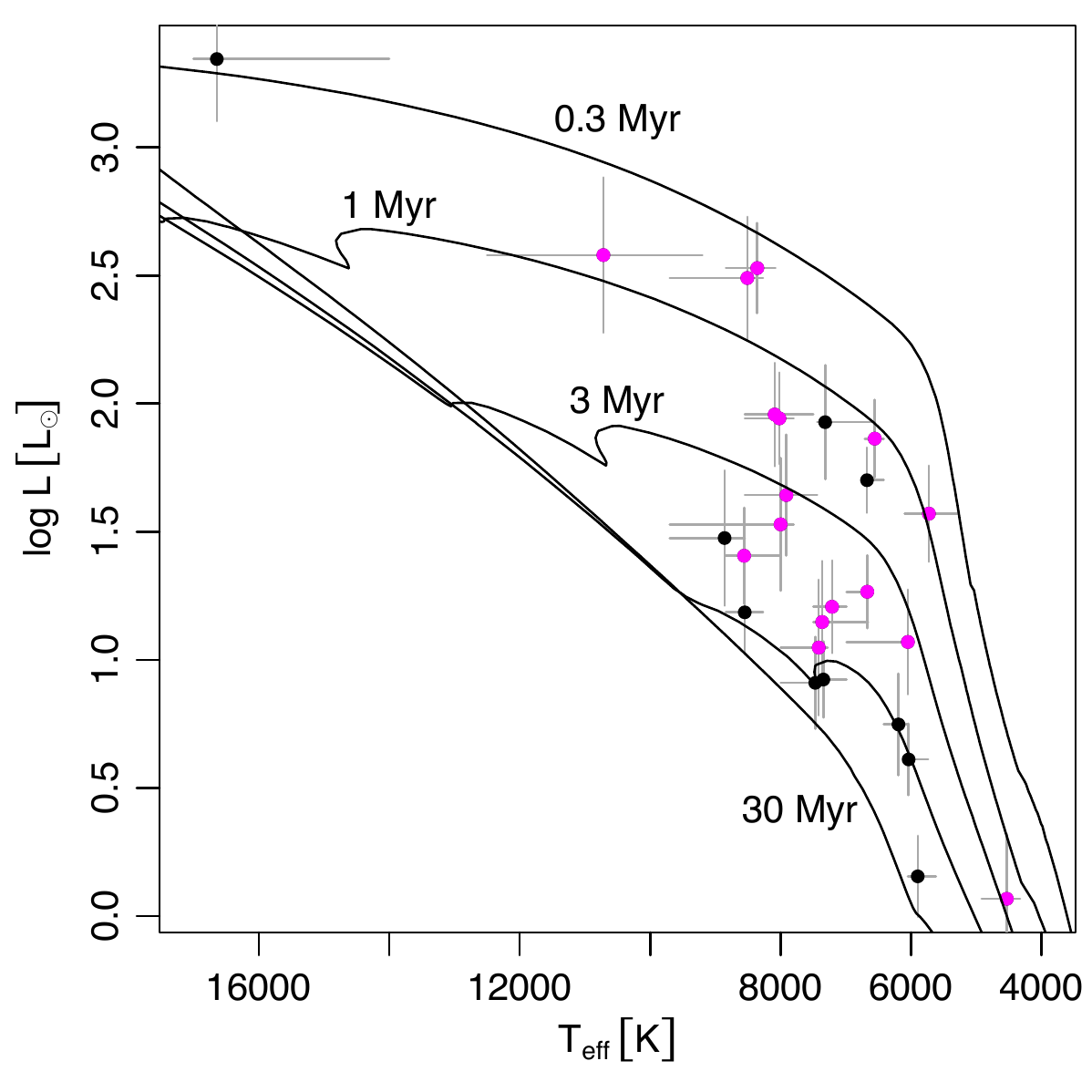}
\caption{HR diagram for the observed SPICY stars.  PARSEC isochrones are drawn at 0.3, 1, 3, 10, and 30~Myr. Stars with H$\alpha$ emission are color-coded magenta. 
\label{fig:hrd}}
\end{figure}

\vspace*{-0.15cm}
\subsection{Spectral Energy Distribution Fitting}\label{sec:sed}

Spectral energy distributions (SEDs) for the program stars (Figure~\ref{fig:sed}) were constructed from photometry obtained from the 
Pan-STARRS \citep{Chambers2016,Flewelling2020}, IPHAS \citep{Barentsen2014}, APASS 9 \citep{Henden2015}, GLIMPSE \citep{2003PASP..115..953B,2009PASP..121..213C}, 2MASS \citep{2006AJ....131.1163S}, WISE \citep{2010AJ....140.1868W}, SDSS DR12 \citep{Alam2015}, Tycho-2 \citep{2000A&A...355L..27H}, VPHAS+ DR2 \citep{2014MNRAS.440.2036D}, DENIS \citep{2000A&AS..141..313F}, and MIPSGAL \citep{2009PASP..121...76C,2015AJ....149...64G} surveys.\footnote{
The survey photometry were queried using the Virtual Observatory SED Analyzer \citep[VOSA;][]{Bayo2008}, but SED fitting was performed in R using the methods described here.} These surveys collectively cover 0.36--24~$\mu$m in wavelength. 

Source properties, including extinction and luminosity, are inferred by modeling the optical and near-infrared portion of the SEDs, where the star (rather than the disk) is expected to dominate the flux. To exclude disk emission, we set a cutoff at the wavelength (somewhere between $H$-band to the $5.8$~$\mu$m band), beyond which, photometric points are consistently above the stellar photosphere by at least two standard deviations. Bands with infrared excess are not included in SED fitting. 

We fit these photometric data points with reddened stellar atmosphere models from 
\citet{CastelliKurucz2003}, which have been convolved with the filter profiles. In these models, we set $T_\mathrm{eff}$ to the value found from spectroscopy but allow extinction, parameterized as $A_V$, to vary. We used the extinction law from \citet{Cardelli1989} with $R_V=3.1$ in the optical/near-infrared and \citet{Wang2015} in the mid-infrared. 
Given that young stars are variable, we used models that include an additional, normally distributed, error term, $\varepsilon$, in fluxes beyond statistical photometric uncertainties. The log-likelihood is then
\begin{equation}
    \ln \mathcal{L} = -\sum_{i=1}^{n_\mathrm{band}} \frac{( y_i - y_\mathrm{model,i})^2}{\sigma_\varepsilon^2 + \sigma_\mathrm{phot}^2} - \frac{n_\mathrm{band}}{2}\ln\left(\sigma_\varepsilon \sqrt{2\pi}\right),
\end{equation}
where $n_\mathrm{band}$ is the number of bands, $y_i$ is the log of the flux in the $i$-th band, $y_{\mathrm{model},i}$ is the log of the model flux in that band, $\sigma_\varepsilon$ is standard deviation of $\varepsilon$, and $\sigma_\mathrm{phot}$ is the photometric uncertainty. Maximum likelihood values of $A_V$, $\sigma_\varepsilon$, and the flux scaling are calculated via the
BFGS algorithm \citep{BFGS_B, BFGS_F, BFGS_G, BFGS_S} implemented in the {\tt optim} function \citep{RCore2021}. Uncertainties on model parameters are determined from the inverse of the Hessian of the log-likelihood at the maximum.

Figure~\ref{fig:sed} shows the stellar SEDs, with the corresponding best-fit reddened stellar atmosphere models overlaid. For all program stars, observed mid-infrared fluxes lie above the stellar atmosphere models, confirming the presence of infrared excesses identified in \citetalias{SPICY}.

\subsection{Extinction}\label{sec:av}

All program stars have moderate to heavy extinction, with values derived from SED fitting ranging from $A_V=1.3$~mag to 5.8~mag, with a median of $A_V=2.3$~mag. The uncertainties on extinction account for both the statistical uncertainties from model fitting and systematic uncertainties due to the choice of $T_\mathrm{eff}$. Typical uncertainties are $\pm$0.2~mag, but range from $\pm$0.1~mag to $\pm$0.6~mag. The sources with the largest uncertainties have the largest scatter in photometric measurements, likely due to variability. 

Diffuse interstellar bands (DIB) are detected in all spectra (Table~\ref{tab:dib}). The strengths of these features are correlated with extinction \citep{2018PASP..130g1001K}, so they can be used to corroborate $A_V$ estimates for YSOs independently from SED fitting \citep{Carvalho2022}. Figure~\ref{fig:dib1} (left) shows the relation of the equivalent width of the 5780~\AA\ DIB and our SED-based $A_V$ estimates.  The \citet{Kendall1948} rank correlation test -- a robust test for statistical correlation -- indicates that the positive correlation between these measures of extinction is statistically significant, with a null-hypothesis probability $p < 10^{-5}$.

The interstellar medium along a particular line of sight may influence the relation between DIB equivalent width and $A_V$, with lines of sight being classified as $\sigma$~Sco-like, lower $W(5780)/W(5797)$ ratios, or $\zeta$~Oph-like, higher $W(5780)/W(5797)$ ratios \citep{Vos2011,2013ApJ...774...72K,2015MNRAS.452.3629L}. We depict both relations on Figure~\ref{fig:dib1}. The points are broadly consistent with these lines, with a slight tendency toward the upper, $\sigma$~Sco-like relation. Overall, the level of scatter is similar to that seen in studies of nearby YSOs \citep{Carvalho2022}.

Contributions to extinction come from the foreground, the local clouds, and circumstellar dust. We used the \citet{2019ApJ...887...93G} extinction map to examine the contribution from the foreground and the local clouds. This map was queried\footnote{\url{http://argonaut.skymaps.info}} to obtain reddening as a function of distance modulus ($DM$) in the direction of every program star on a grid with bin spacings of 0.125 mag. Reddening from the map  was then converted to extinction assuming $A_V = 3.04\,E(g-r)$.
In Figure~\ref{fig:dib1} (right), the ordinate is the line-of-sight extinction from the map, measured to $DM-0.2$ (in front of the star) and to $DM+0.2$ (behind the star), bracketing the YSO and its local cloud. The abscissa is the $A_V$ from our SED fitting. If there is a jump between $DM-0.2$ and $DM+0.2$, it suggests contribution from the local cloud.

The map-based $A_V$ values are correlated with the SED-based $A_V$ values (Figure~\ref{fig:dib1}, right). However, at $DM-0.2$, the map only accounts for $\sim$80\% (median) of the observed extinction, increasing to $\sim$90\% (median) at $DM+0.2$. 
This suggests that circumstellar dust, which is not accounted for in the map, may also contribute to the observed $A_V$. Only five stars have jumps in extinction greater than half a magnitude between $DM-0.2$ and $DM+0.2$, implying that the extinction is largely foreground for most stars, i.e., these SPICY stars are not deeply embedded.

\subsection{Luminosities}\label{sec:lum}

Photospheric bolometric luminosities ($L_\mathrm{bol}$) are derived from the Gaia EDR3 $G_{RP}$ magnitudes, with bolometric corrections ($BC$) and extinction corrections ($A_{RP}$) calculated from pre-main-sequence models. The mean photometry provided by Gaia reduces scatter from stellar variability, and the $RP$ band is less susceptible to either accretion luminosity or variations in extinction law than the $BP$ band. $BC$ and $A_{RP}$ were calculated as functions of $T_\mathrm{eff}$ and $A_V$ by interpolating over values from the \citet{Bressan2012} isochrone tables to account for nonlinearities in reddening relations as a consequence of the breadth of the Gaia bands. The bolometric luminosity, with all dependencies, is 
\begin{multline}
\log L_\mathrm{bol} = -\mathlarger{[}RP - 5\,\log(\varpi) + 10 \\
- A_{RP}(A_V,T_\mathrm{eff}) +  BC(T_\mathrm{eff}) - 4.74\mathlarger{]} \big/2.5,
\end{multline}
where $L_\mathrm{bol}$ has units of $L_\odot$ and $\varpi$ has units of mas. Note that $L_\mathrm{bol}$ is the stellar bolometric luminosity, excluding any additional luminosity derived from accretion that may contribute to ultraviolet and infrared excesses.
    
\subsection{Stellar Masses and Ages}\label{sec:mass}

Assuming that the program stars are pre-main-sequence, then stellar ages and masses can be estimated from the Hertzsprung-Russell (HR) diagram. However, different theoretical evolutionary models produce age and mass estimates with substantial systematic differences \citep[][]{David2019,Braun2021}. Here, we assume the models from PARSEC version 1.2S \citep{Bressan2012,Chen2014,Chen2015}. 

Figure~\ref{fig:hrd} shows the program stars on the HR diagram, with PARSEC tracks between 0.3~Myr and 30~Myr indicated. All program stars are within the region of the diagram consistent with the pre-main sequence. Stellar masses estimated from the PARSEC models range from 1--6.6~$M_\odot$, with a median of 2~$M_\odot$, and ages range from 0.3--26~Myr, with a median of 4.4~Myr. 

In our sample, 12 out of 15 stars younger than 5~Myr have H$\alpha$ emission, whereas only 4 out of 11 stars older than 5~Myr do. Furthermore, no star with an age estimate $>$10 Myr has H$\alpha$ emission.

\subsection{Accretion Rates}

Accretion luminosity for YSOs is mostly emitted in the ultraviolet \citep{2016ARA&A..54..135H}, but a variety of optical emission lines have been found to strongly correlate with accretion luminosity and can be used to estimate accretion rates \citep[e.g.,][]{Ingleby_2013,Wichittanakom2020}. 

We convert the H$\alpha$ equivalent widths to H$\alpha$ luminosities by multiplying by the continuum from the dereddened stellar model (Section~\ref{sec:sed}) and scaling for distance. We assume the empirical relation between accretion luminosity $L_\mathrm{acc}$ and H$\alpha$ luminosity found by \citet{2017MNRAS.464.4721F} for Herbig Ae/Be stars,
\begin{equation}
\log (L_\mathrm{acc}/L_\odot) \approx A + B\,\log(L_\mathrm{H_\alpha}/L_\odot),
\end{equation}
where $A = 2.09 \pm 0.06$ and $B=1.00\pm0.05$. Then, estimate accretion rate $\dot{M}$ with the equation
\begin{equation}
    \dot{M} \approx \frac{L_\mathrm{acc} R_*}{G M_*},
\end{equation}
where $G$ is the gravitational constant, $M_*$ is stellar mass, and $R_*$ is stellar radius \citep{Calvet1998}. Differences in accretion mechanism may be a source of systematic error, since T Tauri stars are expected to have magnetospheric accretion (for which the equation applies), whereas Herbig Ae/Be stars are likely to undergo boundary layer accretion.

For the 16 stars with measured $W(\mathrm{H}\alpha)$, accretion rates range from $\dot{M}=3\times10^{-8}$--$3\times10^{-7}$~$M_\odot$~yr$^{-1}$, with a median value of $7\times10^{-8}$~yr$^{-1}$ (Table~\ref{tab:ew}). These rates match the expectations for accretion rates for young stellar objects in this mass range \citep{2015MNRAS.453..976F,Wichittanakom2020,Vioque2022}. For the 16 accreting stars, we find no statistically significant correlation between $\dot{M}$ and either pre-main-sequence age or spectral index of the infrared excess. However, our statistical ability to probe these relations may be limited by our small sample size. 

\section{Validation of the SPICY Catalog}\label{sec:validation}

\subsection{Classification as Pre-Main-Sequence Stars}

All 26 program stars lie above the main sequence, with temperatures and luminosities consistent with expectations for pre-main-sequence stars. This evidence, combined with the detection of significant infrared excess from all sources and strong emission lines from more than half the sources, strongly indicates the youth of this sample.

In the mid-infrared, where SPICY selection was performed, the objects whose colors are most easily confused with YSOs are highly-reddened background evolved stars \citep{2011ApJS..194...14P,Povich2013}, including asymptotic-giant-branch (AGB) stars and post-AGB stars \citep{2008AJ....136.2413R,2020ApJ...891...43S}, dusty red-giant-branch (RGB) stars \citep{2012A&A...540A..32G}, post-RGB stars produced by binary interaction \citep{2016A&A...586L...5K}, and compact planetary nebulae (PNe) \citep{2010ApJS..186..259R}. 
Classical Be stars with small infrared excesses from circumstellar free-free emission \citep{2010ApJS..186..259R,Rivinius2013}, symbiotic stars \citep{WatersWaelkens1998}, and cataclysmic variables (CVs) with infrared excess \citep{2004MNRAS.349..869D,2019MNRAS.483.5077A} are other possible contaminants. Extragalactic sources, predominantly starburst galaxies and obscured active galactic nuclei \citep{2005ApJ...631..163S,2009ApJS..184...18G,2011ApJ...735..112J} are well-known contaminants for deeper surveys of star-forming regions at higher Galactic latitudes but are generally too faint to constitute a significant contaminating population for SPICY.
Each of these categories can be ruled out for our program stars. 

The SED analysis shows that all program stars have robust infrared excesses, implying that the infrared colors are not merely the result of high reddening. The positions on the HR diagram rule out AGB and post-AGB stars. AGB stars have $T_\mathrm{eff} < 5000$~K and $L>10^2$~$L_\odot$ \citep{2013MNRAS.434..488M}, which does not match any star from our sample. In
the temperature range of the program stars, the lowest luminosity on the post-AGB is $\sim$2$.5\times 10^3$~$L_\odot$ \citep{2016A&A...588A..25M}, which is $\sim$0.7~dex more luminous than the brightest object from our sample. The optical spectra of PNe have minimal continuum components and are easily distinguished from stars.
RGB stars with mass loss have spectral types of K or M, and there are very few examples with luminosities below 600~$L_\odot$, ruling out this category. CVs have broad emission lines and smooth continua, easily distinguished from YSOs. Based on the Classical Be sample from \citet{Vioque2020}, which we have cross-matched with Spitzer photometry, Classical Be stars rarely have spectral indices exceeding $-1.5$ in the mid-infrared. In contrast, in our sample, the B4 star has $\alpha = -1$ and the B9e star has $-1.4$. Our spectral classification can rule out symbiotic stars containing a red giant. However, as noted by \citet{WatersWaelkens1998}, this category can be tricky to distinguish from Herbig Ae/Be stars entirely. Post-RGB stars are expected to have $0<\log g<2$ \citep{2016A&A...586L...5K}, whereas our spectra are all consistent with $\log g > 2$ (Section~\ref{sec:spectra}). Finally, extragalactic contaminants were avoided by our parallax criterion (Section~\ref{sec:sample}). 

\citetalias{SPICY} reclassified many ``candidate AGB stars''  from \citet{2008AJ....136.2413R} to ``candidate YSOs,'' four of which are program stars confirmed here: SPICY~93027, SPICY~103533, SPICY~105733, and SPICY~106859. Our spectral types for these stars are A3, F3, A6, and A9. These results show the value of selecting YSO candidates based on complex multidimensional color criteria, accomplished in \citetalias{SPICY} with the use of the random-forest algorithm, rather than heuristic cuts on color-color diagrams. 

\subsection{An Upper Limit on Contamination Rate}

The confirmation of 26 out of 26 program stars as pre-main-sequence stars suggests that the contamination rate is low among SPICY sources with similar properties to those we observed.
Given our results, an upper limit on contamination rate may be calculated using Bayes' theorem. Assuming that the number of contaminants, $k$, in a sample of $N$ stars, follows the binomial distribution, with a contamination rate, $r$, that has a uniform prior between 0 and 1, then the posterior distribution for contamination rate is the beta distribution \citep{Bayes1763}
\begin{equation}
    p(r|k,N) = \frac{
    (N+1)!}{k!(N-k)!}r^k(1-r)^{N-k}.
\end{equation}
For the distribution given by $N=26$ and $k=0$, there is a 95\% probability that $r < 10\%$.  

Strictly speaking, this contamination rate only applies to SPICY sources with similar characteristics as our sample (Section~\ref{sec:sample}). The isolation of the observed stars would, ostensibly, increase the odds of them being contaminants, but it is less intuitively evident how the other selection requirements (visual brightness, astrometric properties) would affect the relative contamination rate.
The sample does not represent SPICY stars with the reddest $H-K$ colors, including stars with high near-infrared extinctions or intrinsically red near-infrared colors. 
Nevertheless, the low contamination rate suggests that the SPICY classification methodology successfully separated out numerous categories of potential contaminants. 

\begin{figure}[t]
\centering
\includegraphics[angle=0.,width=0.5\textwidth]{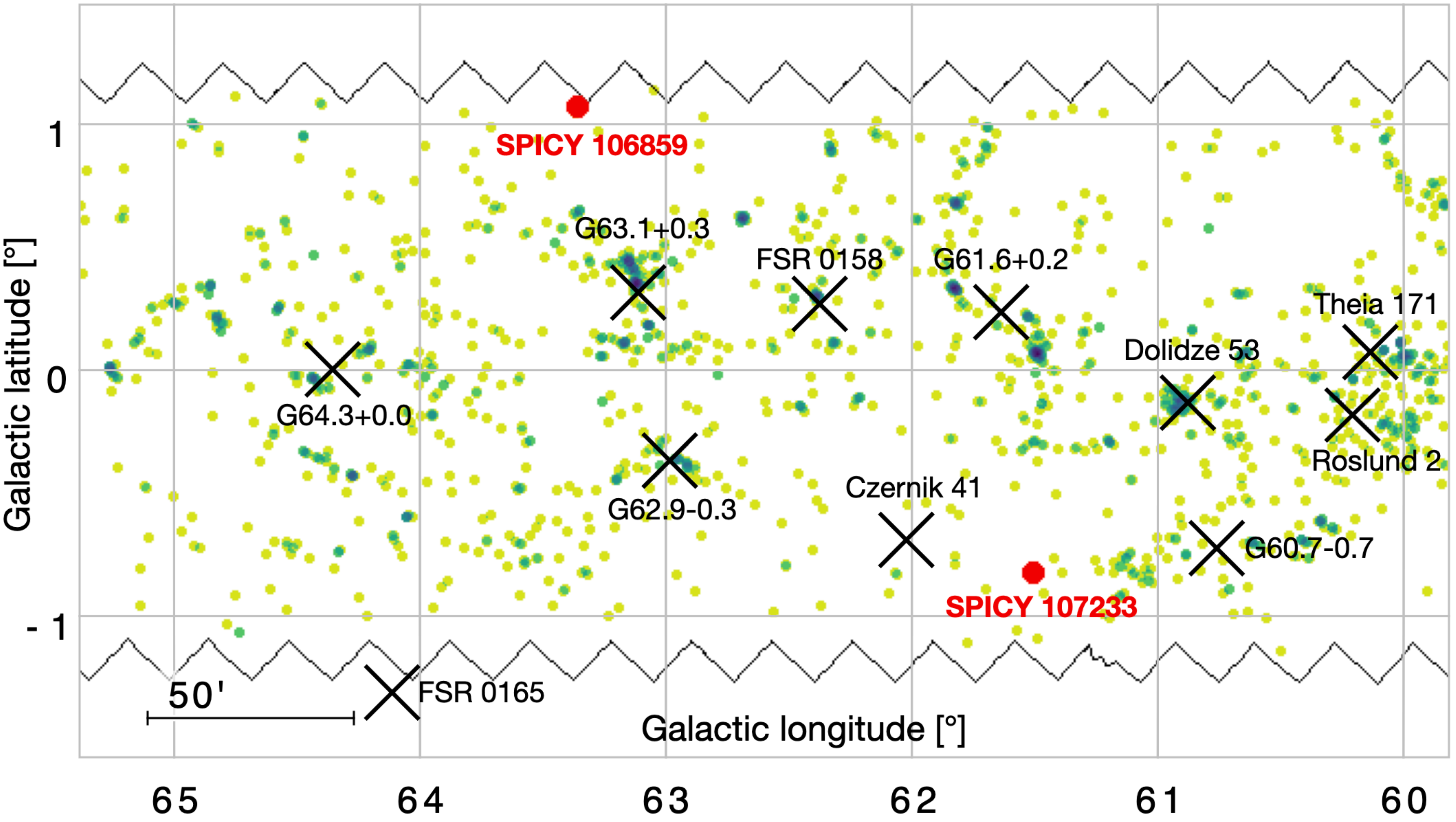}
\caption{A representative section of the GLIMPSE survey area (boundary shown in black) containing two of the program stars (red circles). Other SPICY stars are indicated by the green points. Young stellar groups from \citetalias{SPICY}, supplemented by groups from \citet{CantatGaudin2020} and \citet{Kounkel2020}, are indicated by the black X's. 
\label{fig:lb}}
\end{figure}

\begin{figure*}[t]
\centering
\includegraphics[angle=0.,width=0.85\textwidth]{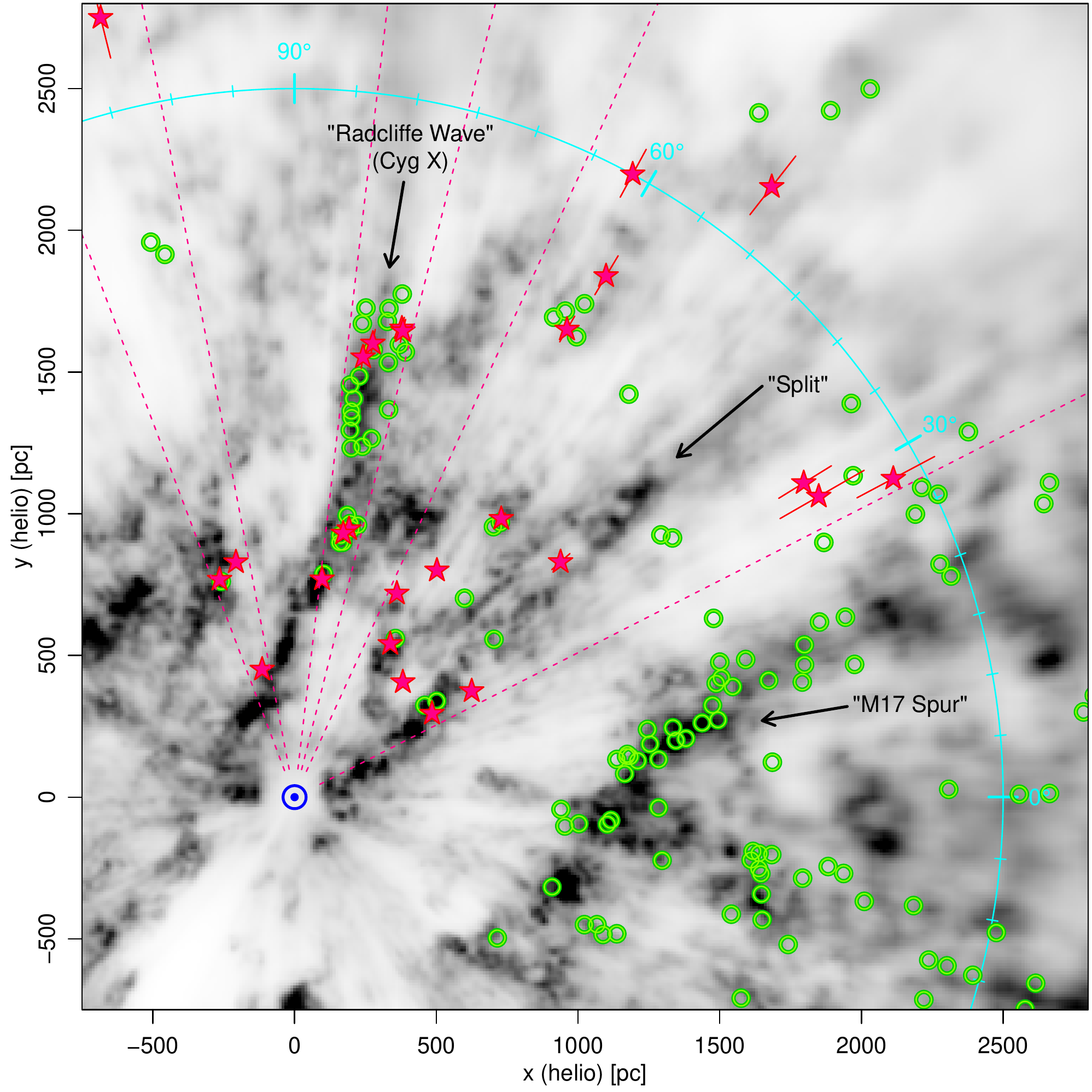}
\caption{Locations of YSO groups (green circles) and isolated YSOs with spectra (magenta stars) plotted on the \citet{Vergely2022} extinction map (grayscale). The map uses a heliocentric Galactic $(x,y)$ coordinate system, with the Sun at $(0,0)$ (blue symbol), the Galactic Center on the positive $x$ axis, and the direction of rotation at the location of the Sun parallel to the $y$ axis. Dashed magenta lines indicate the boundaries (in Galactic longitude) of the area from which the program stars were selected.   
\label{fig:galmap}}
\end{figure*}

\section{Galactic Environment}\label{sec:isolation}

Given the high density of young stellar clusters in the Galactic midplane (e.g., $\sim$400 SPICY groups within a 613 square-degree survey area), even the most isolated  YSOs tend to be separated, in projection, by only a degree or so from the nearest star-forming regions (e.g., Figure~\ref{fig:lb}). Thus, a 3D perspective is needed to determine whether the program stars are on the outskirts of star-forming regions or truly on their own.

Gaia's astrometry provides this 3D view of structures in the Solar neighborhood. Figure~\ref{fig:galmap} shows the positions of the program stars on a Gaia-based map of dust within $\sim$3~kpc of the Sun \citep{Vergely2022}. The map uses heliocentric $(x,y)$ positions, oriented with the Galactic Center on the positive $x$-axis and the direction of Galactic rotation parallel to the $y$-axis. Program-star distances are computed from the reciprocal of the zero-point corrected parallax, and asymmetric 1$\sigma$ error bars are shown. The map also includes the YSO groups from \citetalias{Kuhn2021_sgr}. The dashed magenta lines demarcate the portion of the Galaxy examined here (i.e., the fields covered by SPICY that were also visible during the Palomar/DBSP observations.)

Previous studies have shown that star-formation in the Solar neighborhood tends to be concentrated in dusty kpc-scale filamentary structures. Several of these, including the Radcliffe Wave \citep{Alves2020,Zucker2020}, the Split \citep{Lallement2019}, and the M17 Spur \citepalias{Kuhn2021_sgr}, have been marked on Figure~\ref{fig:galmap}. Both the Radcliffe Wave (the Cyg~X section) and the M17 Spur contain dozens of SPICY YSO groups. The Split only contains a few YSO groups with Gaia distance estimates, but this deficit in the ``Split'' can be partially attributed to high extinction. Many of the program stars also reside near the filamentary cloud structures that appear in the \citet{Vergely2022} maps, including the Radcliffe Wave and the Split (the M17 Spur was not visible from Palomar at the time of observation).

In our investigation of spatial distribution, we are interested in determining 1) whether the program stars are associated with dusty Galactic structures and 2) whether the program stars are preferentially near groups of other YSOs. The statistical significance of these can be assessed via simulations. For each simulation, we generated 26 artificial stars with the same parallaxes as the observed stars but positioned randomly in $(\ell,b)$ throughout the survey area. We used 10,000 simulation realizations and computed the same statistics for both the observed and simulated stars. The statistics used for the tests are 1) the median value of the extinction map at the locations of the stars and 2) the median distance from each star to the nearest YSO group. Both tests reveal moderate statistical significance ($p<0.05$), indicating that the program stars are preferentially located nearer dusty structures and YSO groups than would be expected for a random spatial distribution. This implies that, even if these stars are not members of discrete clusters, they are likely connected to the same star-forming events that produced these large-scale structures.

In several cases, the program stars may be outlying cluster members not identified by the \texttt{HDBSCAN} algorithm. To examine this possibility, we searched a compilation of young clusters and associations from \citet{CantatGaudin2020},  \citet{Kounkel2020}, and \citetalias{SPICY}, including only clusters with estimated ages $<$30~Myr for the former two catalogs. Six program stars are separated, in projection, by $s<10$~pc from nearby young clusters that have statistically identical parallaxes and tangential velocities differing by $<$3~km~s$^{-1}$. These star/cluster pairs include: SPICY~103533 and G53.1+0.0 ($s = 5$~pc), SPICY~105589 and G57.5+0.2 ($s = 6$~pc), SPICY~108400 and Berkeley~86 ($s = 9$), SPICY~11327 and G79.4-0.5 ($s = 8$~pc), SPICY~115897 and G82.5+0.1 ($s = 3$~pc), and SPICY~117231 and G108.7+2.3 ($s = 8$~pc). In addition, there are seven stars whose projected trajectories would intersect known clusters on a timescale consistent with their ages, meaning that some of these could have been ejected. However, the crowding of clusters means that these solutions are not unique, so we do not provide them here. 

Finally, to check whether the program stars have siblings not identified via infrared-excess, we examined Gaia sources in a $0.5^\circ$ radius around each program star to determine whether there are stellar overdensities in the vicinity of the star corresponding to clusters that had not been previously noted. We performed ADQL queries to identify all Gaia EDR3 sources with parallaxes consistent with the SPICY stars within 1$\sigma$ Gaia parallax uncertainties and proper motions that differ by $\leq$1~mas~yr$^{-1}$. We further required the $\mathtt{astrometric\_sigma5d\_max} < 0.5$~mas, $\mathtt{RUWE}<1.4$, and  $\mathtt{astrometric\_exces\_noise}<$1~mas. We expect unrelated field stars selected by chance to be distributed with complete spatial randomness within the circular query region. We found that SPICY~103533, SPICY~104101, SPICY~105586, SPICY~108560, SPICY~111557, and SPICY~111583 had inhomogeneities in the spatial distribution of nearby Gaia sources, but no distinct cluster was found. 
Overall, the results from the spatial analysis suggest that the isolated YSOs tend to be peripheral members of larger-scale star-forming Galactic structures rather than the product of star-formation in complete isolation. 

\section{Discussion and Conclusion}\label{sec:dc}

The SPICY catalog is one of the most extensive compilations of YSO candidates, with 117,446 total sources. The catalog, based on Spitzer's GLIMPSE and related surveys, covers most of the Galactic midplane in the inner Galaxy, where numerous star-forming regions are found. Although many of the YSO candidates are members of groups -- some of which were first identified in \citetalias{SPICY} -- a surprisingly high fraction do not appear clustered (Section~\ref{sec:definition}), leaving no portion of the survey area completely devoid of YSO candidates. 

We have spectroscopically examined  26 of the ``isolated'' candidates to determine whether they are {\it bona fide} YSOs. The results, including their location in the pre-main-sequence part of the HR diagram, emission lines (H and Ca\,{\sc ii}) from over half the stars, and confirmation of the presence of infrared excess, demonstrate that all are YSOs. Spectral types range from B4 to K3, with spectral types of A being the most common. It is likely that the relatively bright limit of $G<15$~mag imposed on the spectroscopic sample skews the distribution toward earlier-type stars, so this spectral type distribution is not likely to be representative of the SPICY catalog as a whole. The stars are moderately to heavily obscured, with extinctions ranging from 1--6~mag in the $V$-band. We have corroborated our extinction estimates using measurements of DIBs and Galactic extinction maps. For stars with H$\alpha$ emission lines, we calculate accretion rates of $\log \dot{M}[M_\odot\,\mathrm{yr}^{-1}]\sim-7.5$ to $-6.5$, which is typical for accreting young stars in this mass range (1--7~$M_\odot$).

These results imply that a sizable fraction (up to $\sim$50\%) of YSOs are located in low-density environments rather than dense groups. Although our spectroscopic sample is fairly small, given the 100\% confirmation rate, we can put an upper limit on the contamination rate of $\sim$10\%, which is small enough that contamination is incapable of having a significant impact on the ratios of clustered to non-clustered YSOs. This contamination rate is only strictly valid for other stars meeting the same selection criteria we used for the program stars. Using our definition of clustered from Section~\ref{sec:definition}, approximately half of these stars are clustered and half isolated. This ratio extends to the rest of the SPICY stars too, albeit the contamination rate for the non-optically visible stars is less certain.

Another caveat is that infrared-excess selection of YSOs does not produce an unbiased representation of the spatial distribution of YSOs in the Galaxy. In particular, within massive star-forming regions, strong crowding and bright mid-infrared nebulosity make it difficult to identify stars based on infrared excess \citep{Kuhn2013,Richert2015}. In such regions, young stars identified by other methods tend to be in the majority \citep{2013ApJS..209...32B}, and this has been shown to skew estimates of YSO surface densities from mid-infrared surveys toward lower values \citep{2015ApJ...802...60K}. Similarly, SPICY is likely to under-count the fraction of YSOs in the richest, most active star-forming regions.

A significant fraction of stars born in low-density environments has implications for the mechanisms of star formation. Molecular clouds with small, self-gravitating regions are expected to lead to low-efficiency, distributed star formation. Meanwhile, more massive clouds with larger self-gravitating regions are expected to form stars more efficiently, producing denser clusters of stars \citep{2017MNRAS.472.4982S, 2018PASP..130g2001G}. In addition, theoretical studies suggest that strong supersonic turbulent support of molecular clouds could be a mechanism that limits the size of self-gravitating clumps, leading to distributed star formation \citep{2004RvMP...76..125M}. Thus, the finding of large numbers of distributed stars suggests that turbulence could be an important regulatory mechanism for star formation in the Solar neighborhood. However, other mechanisms can also produce YSOs in relative isolation. For example, if a star-forming molecular cloud hub is being fed by filaments, and the system is disrupted by ionizing radiation, cutting off the filaments, then the filaments may leave behind trails of isolated star-forming cores \citep{2019MNRAS.490.3061V}.

In the Solar neighborhood, the clustered YSO populations tend to be concentrated in kpc-long dusty filaments, including the Radcliffe Wave \citep{Alves2020}, the Split \citep{Lallement2019}, and the M17 Spur \citepalias{Kuhn2021_sgr}. Three-dimensional analysis of the isolated YSO distribution suggests that these stars, as well, are preferentially located in the vicinity of these structures. These results suggest that the processes that shape these filaments are important in facilitating star formation on Galactic scales.

Several YSOs in this study meet the definition for Herbig Ae/Be stars suggested by \citet{WatersWaelkens1998}, including 1) a spectral type of A or B, 2) emission lines, 3) infrared excess from dust, and 4) a luminosity class of III--V. The program stars meeting these criteria include the Be star SPICY~111583, and 9 Ae stars, SPICY~93027, 98558, 104101, 105733, 106859, 108375, 108400, 111557, and 113327. Herbig stars are scientifically interesting as they bridge a gap between high and low-mass star formation, but confirmed Herbig stars have, historically, been relatively rare \citep{WatersWaelkens1998}. Nevertheless, this is beginning to change, with 128 new Herbig stars from \citet{Vioque2022} and the promise of many new Herbig stars from Gaia's third data release.

\appendix

\section{Notes on Individual Stars}\label{appendix:individual}

Three of the program stars were classified by \citet{2008AJ....136.2413R} as candidate YSOs (SPICY~89018, SPICY~90923, and SPICY 98558), and four were classified as candidate AGB stars (SPICY~93027, SPICY~103533, SPICY~105733, and SPICY 106482). 

For the readers' convenience, in the section headers for the stars discussed below, we provide spectral types and distance from Tables~\ref{tab:log} and \ref{tab:spec}.

\subsection{SPICY 89954 (G0; 730~pc)}

In 2MASS, the source is blended with a nearby star.

\subsection{SPICY 90923 (F0; 570~pc)}

There are 10 Gaia sources within $0.5^\circ$ of SPICY~90923 with consistent parallaxes and proper motions. Of these ten sources, eight appear randomly scattered throughout the field, but 2 of them (Gaia~DR3~4260271558057495680 and Gaia~DR3~4260271553758475776) are located within 2$^\prime$ of SPICY~90923. The probability of this arrangement occurring by chance is $p<10^{-4}$, suggesting that these stars are physically associated. 

\subsection{SPICY 98558 (A3; 1250~pc)}

This star was classified as a variable star by \citet{2018AJ....156..241H}. There is a small patch of mid-infrared nebulosity $\sim$6$^{\prime}$ to the northwest. This patch, HRDS~G041.515-00.139 ($=$~IRAS 19052+0729), is $\sim$1$^\prime$ in diameter, but the nebulosity does not extend to the location of SPICY~98558. 

The spectrum shows N\,{\sc i} absorption lines around 8700--8720~\AA, which are typically more prominent in lower-surface-gravity A-type stars \citep{2009ssc..book.....G}. However, these have also been detected in the spectra of the YSO RNO~1 \citep{Carvalho2022}.

\subsection{SPICY 106859 (A9; 805~pc)}

This is the variable star IZ Vul, identified by \citet{1968IBVS..311....1K}. \citet{Vioque2022} assigned the star a spectral type of A2 (VOS~104 in their catalog).
Our preferred spectral type, A9, provides a better fit to features in the DBSP spectrum, including the Balmer line profiles and the strength of the Ca~K line, which is significantly underestimated if a spectral type of A2 is assumed.  

\subsection{SPICY 108560 (F0; 970~pc)}

This star was classified as cluster member NGC 6910 SIC 28 by \citet{1991AZh....68..466S}. However, the parallax of the cluster ($0.545\pm0.004$~mas) and parallax of the star ($1.04\pm0.01$~mas) are not compatible. \citet{Vioque2022} determined a spectral type of A8 for this star, which is within the confidence interval of our preferred spectral type F0$^{-2}_{+1}$.  

\subsection{SPICY 113327 (A9; 950~pc)}

This star, V1394 Cyg, was classified as a candidate LPV star  \citep{1969MmSAI..40..375R}.

\subsection{SPICY 115897 (A2; 780~pc)}

This star, TYC 3174-358-1, was examined by \citet{2018ApJ...868...43S} who argue that the WISE 22~$\mu$m excess is associated with a background object rather than the star itself.

\subsection{SPICY 116344 (K3; 465~pc)}

This star was identified as an emission line star by \citet{1999A&AS..134..255K}. 

\subsection{SPICY 117231 (F7; 810~pc)}

This star is IRAS 22489+6107. The spectral type A9 from \citet{Vioque2022} overestimates both the line widths and strengths of the H$\gamma$ and H$\delta$ lines in our DBSP spectrum. Thus we prefer a later spectral type of F7. 

%\begin{acknowledgments}
\vspace{2mm}
\noindent R.S.\ was supported by Caltech's Freshman Summer Research Institute (FSRI). We would like to thank Christoffer Fremling and Milan Roberson for help with DBSP software, Gregory Herczeg for assistance with observations, Rosine Lallement for access to extinction maps, and Adolfo Carvalho for valuable discussions about DIBs. We would like to thank the referee for providing a thorough report and many useful suggestions. AKM acknowledges the support from the Portuguese Funda\c c\~ao para a Ci\^encia e a Tecnologia (FCT) through grants PTDC/FIS-AST/31546/2017, UID/FIS/00099/2019 and EXPL/FIS-AST/1368/2021. This work is based, in part, on data from ESA's Gaia mission \citep{2016A&A...595A...1G}, processed by the Data Processing and Analysis Consortium, funded by national institutions, particularly those participating in the Gaia Multilateral Agreement. This work also is based, in part, on data from the Spitzer Space Telescope, which was operated by the Jet Propulsion Laboratory, California Institute of Technology, under a contract with NASA. The Cosmostatistics Initiative (COIN, \url{https://cosmostatistics-initiative.org/}) is an international network of researchers whose goal is to foster interdisciplinarity inspired by Astronomy.
%\end{acknowledgments}

\facilities{Hale, Gaia, Spitzer (IRAC)}

\software{astropy \citep{2013A&A...558A..33A,2018AJ....156..123A},  
          DBSP\_DRP \citep{2021ascl.soft08020R},
          FITSio \citep{FITSio},
          MASS \citep{MASS},
          R \citep{RCore2021}, 
          TOPCAT \& STILTS \citep{Taylor2005}
          }

\vspace*{-0.4cm}
\bibliography{ms.bbl}{}
\bibliographystyle{aasjournal}

\vspace*{-0.4cm}
\begin{deluxetable*}{lrrrrrrrrrrrrD}[ht]
\tablecaption{SPICY Stars Observed by DBSP \label{tab:log}}
\tabletypesize{\small}\tablewidth{3pt}\rotate
\tablehead{
  &&&&&&\multicolumn{6}{c}{Gaia EDR3} \\
  \cline{7-13}
  \colhead{Name} & \colhead{R.A.} & \colhead{Decl.} & \colhead{Exp.}   &  \colhead{Air.} & \colhead{$\alpha$} & \colhead{G} &\colhead{$G_{BP}$} &\colhead{$G_{RP}$} &\colhead{$\varpi$} & \colhead{$\mu_{\ell^\star}$} & \colhead{$\mu_b$} & \colhead{$d$}\\
    \colhead{} & \colhead{ICRS} & \colhead{ICRS} & \colhead{s}   &  \colhead{} & \colhead{} & \colhead{mag} &\colhead{mag} &\colhead{mag} &\colhead{mas} & \colhead{mas~yr$^{-1}$} &  \colhead{mas~yr$^{-1}$} & \colhead{pc}\\
    \colhead{(1)} & \colhead{(2)} & \colhead{(3)} & \colhead{(4)}   &  \colhead{(5)} & \colhead{(6)} & \colhead{(7)} &\colhead{(8)} &\colhead{(9)} &\colhead{(10)} & \colhead{(11)} & \colhead{(12)} & \colhead{(13)}
}
\startdata
  SPICY 89018   & 18:39:14.05  & $-$03:57:10.3  & 600     & 2.1  &   $-0.3$ &	14.4  &  15.4  &  13.5  &   $0.417\pm0.027$  &   $-0.742\pm0.0269$	&	$-0.134\pm0.027$ & $2393^{+165}_{-145}$ \\
  SPICY 89954   & 18:42:17.96  & $-$01:03:41.0  & 300     & 1.2  &   $-1.2$ & 	14.4  &  15.2  &  13.4  &   $1.372\pm0.027$  &   $-7.746\pm0.0232$	&	$-5.212\pm0.022$ & $729^{+15}_{-14}$ \\
  SPICY 90918   & 18:45:04.61  & $-$00:35:08.1  & 600     & 1.8  &   $-1.0$ & 	14.5  &  15.4  &  13.5  &   $0.473\pm0.024$  &   $-2.207\pm0.0223$	&	$-0.991\pm0.023$ & $2112^{+115}_{-104}$ \\
  SPICY 90923   & 18:45:05.91  & $-$00:59:01.2  & 180     & 1.2  &   $0.1$  &	12.4  &  12.8  &  11.8  &   $1.761\pm0.014$  &   $-7.209\pm0.0137$	&	$-5.599\pm0.014$ & $568^{+5}_{-5}$ \\
  SPICY 93027   & 18:50:18.19  & $-$03:19:09.2  & 300     & 1.2  &   $-1.0$ &	14.8  &  16.2  &  13.6  &   $0.468\pm0.037$  &   $-4.339\pm0.0315$	&	$-0.651\pm0.034$ & $2133^{+185}_{-158}$ \\
  SPICY 98558   & 19:08:01.93  & $+$07:30:01.2  & 600     & 1.4  &   $-0.2$ &	14.6  &  15.5  &  13.7  &   $0.798\pm0.027$  &   $-8.914\pm0.0230$	&	 $0.689\pm0.024$ & $1253^{+44}_{-41}$ \\
  SPICY 100086  & 19:15:35.77  & $+$12:28:13.5  & 600     & 1.5  &   $-1.1$ & 	14.4  &  15.1  &  13.6  &   $1.793\pm0.021$  &   $-3.487\pm0.0167$	&	$-1.702\pm0.016$ & $558^{+7}_{-7}$ \\
  SPICY 103533  & 19:29:38.28  & $+$18:12:40.8  & 600     & 1.4  &   $-0.8$ &	13.8  &  14.4  &  12.9  &   $0.817\pm0.016$  &   $-5.559\pm0.0161$	&	$-1.616\pm0.015$ & $1224^{+25}_{-24}$ \\
  SPICY 104101  & 19:31:13.08  & $+$16:22:03.1  & 300     & 1.3  &   $-0.9$ &	14.1  &  15.1  &  13.1  &   $0.365\pm0.017$  &   $-4.059\pm0.0156$	&	$-0.584\pm0.015$ & $2734^{+138}_{-125}$ \\
  SPICY 105586  & 19:39:27.73  & $+$22:14:26.5  & 180     & 1.2  &   $-1.0$ &	12.5  &  13.0  &  11.8  &   $1.568\pm0.010$  &   $-5.331\pm0.0103$	&	$-3.516\pm0.009$ & $638^{+4}_{-4}$ \\
  SPICY 105733  & 19:40:42.14  & $+$23:18:48.2  & 180     & 1.2  &   $-1.4$ & 13.6  &  14.1  &  12.8  &   $0.466\pm0.017$  &   $-5.832\pm0.0149$	&	$-1.047\pm0.012$ & $2142^{+83}_{-77}$ \\
  SPICY 105879  & 19:41:48.42  & $+$21:44:31.5  & 180     & 1.1  &   $0.5$  &	13.2  &  13.6  &  12.7  &   $1.058\pm0.014$  &   $-4.522\pm0.0149$	&	$-2.403\pm0.015$ & $945^{+14}_{-13}$ \\
  SPICY 106482  & 19:44:45.87  & $+$23:31:46.4  & 180     & 1.1  &   $-1.4$ & 	13.6  &  14.3  &  12.8  &   $0.523\pm0.014$  &   $-5.028\pm0.0121$	&	$-1.447\pm0.010$ & $1910^{+54}_{-51}$ \\
  SPICY 106859  & 19:47:16.49  & $+$27:19:55.7  & 180     & 1.1  &   $-0.5$ & 	12.7  &  13.2  &  12.1  &   $1.242\pm0.013$  &   $-4.373\pm0.0121$	&	$-2.678\pm0.011$ & $805^{+9}_{-8}$ \\
  SPICY 107233  & 19:50:21.04  & $+$24:46:51.5  & 300     & 1.1  &   $-1.0$ & 	14.0  &  15.0  &  13.0  &   $0.399\pm0.015$  &   $-4.992\pm0.0129$	&	$-0.424\pm0.011$ & $2501^{+100}_{-92}$ \\
  SPICY 108375  & 20:20:39.65  & $+$39:12:14.2  & 45     & 1.0  &   $-1.0$ & 	13.0  &  13.5  &  12.3  &   $0.589\pm0.013$  &   $-5.423\pm0.0155$	&	 $0.363\pm0.014$ & $1698^{+40}_{-38}$ \\
  SPICY 108400  & 20:20:45.97  & $+$38:57:01.8  & 300     & 1.1  &   $-0.4$ &	14.0  &  14.5  &  13.3  &   $0.592\pm0.014$  &   $-6.214\pm0.0161$	&	$-0.078\pm0.014$ & $1689^{+43}_{-41}$ \\
  SPICY 108560  & 20:21:36.07  & $+$40:47:56.2  & 90     & 1.0  &   $0.1$  &	12.5  &  12.8  &  12.1  &   $1.035\pm0.010$  &   $-6.723\pm0.0131$	&	$-1.513\pm0.012$ & $966^{+10}_{-10}$ \\
  SPICY 111557  & 20:31:26.94  & $+$41:32:58.3  & 180     & 1.0  &   $-0.8$ &	14.7  &  15.7  &  13.8  &   $0.615\pm0.018$  &   $-4.710\pm0.0218$	&	$-0.325\pm0.021$ & $1625^{+50}_{-47}$ \\
  SPICY 111583  & 20:31:30.94  & $+$42:42:53.3  & 180     & 1.0  &   $-1.4$ & 	12.9  &  13.7  &  12.0  &   $0.636\pm0.016$  &   $-5.169\pm0.0196$	&	 $2.039\pm0.018$ & $1572^{+43}_{-40}$ \\
  SPICY 113327  & 20:35:03.17  & $+$40:13:34.1  & 300     & 1.0  &   $-0.9$ &	14.3  &  15.1  &  13.4  &   $1.056\pm0.015$  &   $-4.114\pm0.0168$	&	$-0.590\pm0.016$ & $947^{+14}_{-14}$ \\
  SPICY 115897  & 20:44:43.73  & $+$42:56:55.3  & 120     & 1.0  &   $1.1$  & 	11.9  &  12.2  &  11.5  &   $1.289\pm0.014$  &   $-2.490\pm0.0160$	&	 $0.041\pm0.015$ & $775^{+9}_{-9}$ \\
  SPICY 116344  & 22:18:54.43  & $+$58:46:57.4  & 300     & 1.2  &   $-0.4$ &	14.3  &  15.1  &  13.3  &   $2.149\pm0.015$  &    $2.897\pm0.0197$	&	$-2.049\pm0.018$ & $465^{+3}_{-3}$ \\
  SPICY 116390  & 22:21:04.95  & $+$57:53:40.2  & 300     & 1.2  &   $-0.5$ &	13.3  &  13.8  &  12.8  &   $1.169\pm0.015$  &   $-1.749\pm0.0155$	&	$-2.312\pm0.014$ & $855^{+11}_{-11}$ \\
  SPICY 116475  & 22:22:27.19  & $+$57:32:21.5  & 300     & 1.2  &   $-1.1$ & 	14.7  &  15.4  &  13.8  &   $0.352\pm0.019$  &   $-4.856\pm0.0210$	&	$-0.981\pm0.020$ & $2834^{+164}_{-147}$ \\
  SPICY 117231  & 22:50:57.07  & $+$61:23:37.7  & 180     & 1.2  &   $0.2$  &	14.2  &  14.9  &  13.3  &   $1.230\pm0.019$  &   $-2.001\pm0.0234$	&	$-2.068\pm0.020$ & $813^{+13}_{-13}$ \\
\enddata
\tablecomments{Sample of 26 SPICY stars observed by DBSP. Column 1: Star name. Columns 2--3: Source coordinates. Column 4: Net exposure time. Column 5: Airmass. Column 6: mid-IR spectral index from \citetalias{SPICY}. Columns 7--13: Gaia EDR3 quantities are provided here for convenience. Parallax (Column 10) includes the zero-point corrections, and distance (Column 13) is the reciprocal of parallax.}
\end{deluxetable*}

\begin{deluxetable*}{lrrrrrrr}[ht]
\tablecaption{Stellar Properties \label{tab:spec}}
\tabletypesize{\small}\tablewidth{3pt}
\tablehead{
  \colhead{Star} & \colhead{SpT} & \colhead{$T_\mathrm{eff}$} & \colhead{$A_V$}   &  \colhead{$\log L$} & \colhead{$R_\star$} & \colhead{age} &\colhead{mass}\\
    \colhead{} & \colhead{} & \colhead{K} & \colhead{mag}   &  \colhead{$L_\odot$} & \colhead{$R_\odot$} & \colhead{Myr} &\colhead{$M_\odot$}\\
    \colhead{(1)} & \colhead{(2)} & \colhead{(3)} & \colhead{(4)}   &  \colhead{(5)} & \colhead{(6)} & \colhead{(7)} &\colhead{(8)} }
\startdata
SPICY 89018 & F4$^{-2}_{+1}$ & 6590$^{+120}_{-170}$ & 3.24$\pm$0.16 & 1.86$\pm$0.15 & 6.5$\pm$1.2 & 1.2$\pm$0.5 & 3.4$\pm$0.5\\
SPICY 89954 & G0$^{-0}_{+3}$ & 6050$^{+60}_{-310}$ & 2.42$\pm$0.16 & 0.61$\pm$0.14 & 1.8$\pm$0.3 & 10.2$\pm$3.7 & 1.5$\pm$0.2\\
SPICY 90918 & F0$^{-1}_{+4}$ & 7280$^{+160}_{-690}$ & 3.80$\pm$0.36 & 1.93$\pm$0.22 & 5.8$\pm$1.6 & 1.4$\pm$1.0 & 3.4$\pm$0.6\\
SPICY 90923 & F0$\pm$1 & 7280$\pm$290 & 1.52$\pm$0.15 & 0.92$\pm$0.15 & 1.8$\pm$0.3 & 10.7$\pm$5.1 & 1.6$\pm$0.1\\
SPICY 93027 & A3$^{-3}_{+1}$ & 8550$^{+1150}_{-280}$ & 5.77$\pm$0.25 & 2.49$\pm$0.24 & 8.0$\pm$2.6 & 0.5$\pm$0.5 & 4.7$\pm$1.0\\
SPICY 98558 & A3$^{-1}_{+3}$ & 8550$^{+290}_{-550}$ & 3.44$\pm$0.18 & 1.41$\pm$0.19 & 2.3$\pm$0.5 & 6.2$\pm$3.0 & 2.0$\pm$0.2\\
SPICY 100086 & G1$^{-1}_{+3}$ & 5970$^{+80}_{-350}$ & 1.74$\pm$0.21 & 0.15$\pm$0.16 & 1.1$\pm$0.2 & 25.7$\pm$6.2 & 1.1$\pm$0.1\\
SPICY 103533 & F3$^{-2}_{+1}$ & 6660$^{+330}_{-70}$ & 2.28$\pm$0.14 & 1.27$\pm$0.14 & 3.2$\pm$0.6 & 4.8$\pm$1.7 & 2.1$\pm$0.3\\
SPICY 104101 & A4$^{-2}_{+1}$ & 8270$^{+570}_{-190}$ & 4.31$\pm$0.15 & 2.53$\pm$0.18 & 8.9$\pm$2.0 & 0.4$\pm$0.3 & 5.0$\pm$0.7\\
SPICY 105586 & F9$^{-8}_{+1}$ & 6090$^{+900}_{-40}$ & 2.02$\pm$0.29 & 1.07$\pm$0.20 & 3.1$\pm$0.9 & 4.3$\pm$3.2 & 2.1$\pm$0.4\\
SPICY 105733 & A6$^{-3}_{+1}$ & 8000$^{+550}_{-200}$ & 2.47$\pm$0.17 & 1.94$\pm$0.18 & 4.9$\pm$1.1 & 1.8$\pm$0.9 & 3.1$\pm$0.5\\
SPICY 105879 & A3$\pm$1 & 8550$\pm$280 & 1.87$\pm$0.15 & 1.19$\pm$0.16 & 1.8$\pm$0.4 & 8.5$\pm$4.1 & 2.0$\pm$0.1\\
SPICY 106482 & F4$\pm$1 & 6590$\pm$170 & 2.20$\pm$0.12 & 1.70$\pm$0.13 & 5.4$\pm$0.8 & 1.7$\pm$0.6 & 3.0$\pm$0.3\\
SPICY 106859 & A9$^{-3}_{+1}$ & 7440$^{+560}_{-160}$ & 1.35$\pm$0.56 & 1.05$\pm$0.26 & 2.0$\pm$0.6 & 9.3$\pm$7.1 & 1.7$\pm$0.2\\
SPICY 107233 & B4$^{-1}_{+3}$ & 16700$^{+300}_{-2700}$ & 4.88$\pm$0.14 & 3.35$\pm$0.24 & 5.6$\pm$1.9 & 0.3$\pm$5.9 & 6.6$\pm$1.1\\
SPICY 108375 & A5$^{-2}_{+3}$ & 8080$^{+470}_{-580}$ & 2.48$\pm$0.22 & 1.96$\pm$0.20 & 4.8$\pm$1.3 & 1.7$\pm$1.0 & 3.2$\pm$0.6\\
SPICY 108400 & A6$^{-6}_{+1}$ & 8000$^{+1700}_{-200}$ & 2.32$\pm$0.30 & 1.53$\pm$0.26 & 3.0$\pm$1.1 & 4.6$\pm$3.0 & 2.2$\pm$0.5\\
SPICY 108560 & F0$^{-2}_{+1}$ & 7280$^{+220}_{-290}$ & 1.31$\pm$0.24 & 1.21$\pm$0.18 & 2.5$\pm$0.6 & 6.9$\pm$2.7 & 1.8$\pm$0.2\\
SPICY 111557 & A6$\pm$3 & 8000$\pm$560 & 3.72$\pm$0.36 & 1.64$\pm$0.24 & 3.4$\pm$1.1 & 3.5$\pm$1.8 & 2.5$\pm$0.5\\
SPICY 111583 & B9$^{-1}_{+2}$ & 10700$^{+1800}_{-1500}$ & 3.78$\pm$0.31 & 2.58$\pm$0.30 & 5.7$\pm$2.6 & 0.8$\pm$1.6 & 4.4$\pm$1.2\\
SPICY 113327 & A9$^{-1}_{+4}$ & 7440$^{+60}_{-780}$ & 3.17$\pm$0.44 & 1.15$\pm$0.24 & 2.2$\pm$0.7 & 8.1$\pm$5.3 & 1.8$\pm$0.3\\
SPICY 115897 & A2$^{-2}_{+1}$ & 8840$^{+860}_{-290}$ & 1.65$\pm$0.45 & 1.48$\pm$0.26 & 2.3$\pm$0.8 & 5.6$\pm$3.6 & 2.1$\pm$0.3\\
SPICY 116344 & K3$^{-2}_{+1}$ & 4550$^{+370}_{-220}$ & 1.74$\pm$0.57 & 0.07$\pm$0.25 & 1.7$\pm$0.6 & 2.5$\pm$5.7 & 1.0$\pm$0.2\\
SPICY 116390 & A9$^{-3}_{+1}$ & 7440$^{+560}_{-160}$ & 1.58$\pm$0.20 & 0.91$\pm$0.18 & 1.7$\pm$0.4 & 10.2$\pm$6.9 & 1.7$\pm$0.1\\
SPICY 116475 & G3$^{-5}_{+4}$ & 5740$^{+360}_{-450}$ & 2.20$\pm$0.26 & 1.57$\pm$0.19 & 6.1$\pm$1.6 & 0.9$\pm$0.8 & 3.4$\pm$0.6\\
SPICY 117231 & F7$^{-2}_{+3}$ & 6140$^{+280}_{-90}$ & 2.40$\pm$0.34 & 0.75$\pm$0.20 & 2.1$\pm$0.5 & 9.1$\pm$4.4 & 1.5$\pm$0.3\\
\enddata
\tablecomments{Stellar properties derived from spectral analysis (Columns 2--3), SED fitting (Columns 4--6), and pre-main-sequence models (Columns 7--8). 
}
\end{deluxetable*}

\begin{deluxetable*}{lrrrrrr}[ht]
\tablecaption{Emission Line Equivalent Widths and Inferred Accretion Rate\label{tab:ew}}
\tabletypesize{\small}\tablewidth{3pt}
\tablehead{
&&&\multicolumn{3}{c}{Ca\,{\sc ii}} &\\
  \cline{4-6}
  \colhead{Star} & \colhead{FW10\%} & \colhead{$W$(H$\alpha$)} & \colhead{$W$(8498)} & \colhead{$W$(8542)}   &  \colhead{$W$(8662)} & \colhead{$\log \dot{M}$}\\
    \colhead{} & \colhead{km\,s{$^{-1}$}} & \colhead{\AA} & \colhead{\AA} & \colhead{\AA}   &  \colhead{\AA} & \colhead{$M_\odot$~yr$^{-1}$} \\
    \colhead{(1)} & \colhead{(2)} & \colhead{(3)} & \colhead{(4)} & \colhead{(5)} & \colhead{(6)}&   }
\startdata
SPICY 89018 & $\ldots$ & $-1.4\pm0.2$ & $\ldots$ &$\ldots$ & $\ldots$& $-7.5\pm0.2$\\
SPICY 89954 & $\ldots$ & $\ldots$ & $\ldots$ & $\ldots$ & $\ldots$ & $\ldots$\\
SPICY 90918 & $\ldots$ & $\ldots$ & $\ldots$ & $\ldots$ & $\ldots$ & $\ldots$\\
SPICY 90923 & $\ldots$ & $\ldots$ & $\ldots$ & $\ldots$ & $\ldots$ & $\ldots$\\
SPICY 93027 & $\ldots$ & $-5.8\pm0.3$ & $-1.6\pm0.1$ & $-2.7\pm0.3$ & $-3.0\pm0.1$ & $-6.6\pm0.2$\\
SPICY 98558 & 590 & $-21.9\pm0.2$ & $-0.8\pm0.1$ & $-1.6\pm0.1$ & $-1.5\pm0.1$ & $-6.5\pm0.1$\\
SPICY 100086 & $\ldots$ & $\ldots$ & $\ldots$ & $\ldots$ & $\ldots$ & $\ldots$\\
SPICY 103533 & 440 & $-7.8\pm0.1$ & $\ldots$ & $\ldots$ & $\ldots$ & $-7.2\pm0.1$\\
SPICY 104101 & $\ldots$ & $-1.6\pm0.1$ & $-1.2\pm0.1$ & $-2.4\pm0.2$ & $-2.2\pm0.1$ & $-7.1\pm0.1$\\
SPICY 105586 & 440 & $-13.2\pm0.1$ & $\ldots$ & $\ldots$ & $\ldots$ & $-7.2\pm0.2$\\
SPICY 105733 & 490 & $-3.0\pm0.1$ & $-0.8\pm0.1$ & $-1.3\pm0.1$ & $\ldots$ & $-7.3\pm0.2$\\
SPICY 105879 & $\ldots$ & $\ldots$ & $\ldots$ & $\ldots$ & $\ldots$ & $\ldots$\\
SPICY 106482 & $\ldots$ & $\ldots$ & $\ldots$ & $\ldots$ & $\ldots$ & $\ldots$\\
SPICY 106859 & 590 & $-26.9\pm0.3$ & $\ldots$ & $\ldots$ & $\ldots$ & $-6.9\pm0.2$\\
SPICY 107233 & $\ldots$ & $\ldots$ & $\ldots$ & $\ldots$ & $\ldots$ & $\ldots$\\
SPICY 108375 & $\ldots$ & $-1.6\pm0.3$ & $\ldots$ & $\ldots$ & $\ldots$ & $-7.5\pm0.2$\\
SPICY 108400 & 830 & $-5.9\pm0.1$ & $-1.9\pm0.1$ & $-2.7\pm0.2$ & $-2.9\pm0.1$ & $-7.2\pm0.2$\\
SPICY 108560 & 490 & $-10.5\pm0.2$ & $-6.9\pm0.1$ & $-7.6\pm1.0$ & $-6.7\pm0.0$ & $-7.1\pm0.2$\\
SPICY 111557 & 1000: & $-5.3\pm0.3$ & $\ldots$ & $\ldots$ & $\ldots$ & $-7.2\pm0.2$\\
SPICY 111583 & 440 & $-7.5\pm0.1$ & $-2.1\pm0.1$ & $-3.8\pm0.3$ & $-3.7\pm0.1$ & $-6.5\pm0.2$\\
SPICY 113327 & 790 & $-12.2\pm0.2$ & $\ldots$ & $\ldots$ & $\ldots$ & $-7.2\pm0.2$\\
SPICY 115897 & $\ldots$ & $\ldots$ & $\ldots$ & $\ldots$ & $\ldots$ & $\ldots$\\
SPICY 116344 & 640 & $-71.9\pm0.2$ & $-1.6\pm0.1$ & $-3.0\pm0.3$ & $-2.6\pm0.1$ & $-7.3\pm0.2$\\
SPICY 116390 & $\ldots$ & $\ldots$ & $-0.3\pm0.1$ & $-0.2\pm0.1$ & $-0.5\pm0.1$ & $\ldots$\\
SPICY 116475 & 540 & $-24.8\pm0.2$ & $-2.6\pm0.1$ & $-5.4\pm0.5$ & $-3.1\pm0.1$ & $-6.6\pm0.2$\\
SPICY 117231 & $\ldots$ & $\ldots$ & $\ldots$ & $\ldots$ & $\ldots$ & $\ldots$\\
\enddata
\tablecomments{Column 2: Full width at 10\% maximum for the H$\alpha$ line. Columns 3--6: Equivalent widths measured as the integrated difference between the best-fitting template and program-star spectra. Negative values indicate emission.}
\end{deluxetable*}

\begin{deluxetable}{lrrrr}[ht]
\tablecaption{DIB Equivalent Widths\label{tab:dib}}
\tabletypesize{\small}\tablewidth{3pt}
\tablehead{
  \colhead{Star} &  \colhead{$W$(4430)} & \colhead{$W$(5780)} & \colhead{$W$(5797)}   &  \colhead{$W$(8620)} \\
    \colhead{} & \colhead{\AA} & \colhead{\AA} & \colhead{\AA}   &  \colhead{\AA} 
}
\startdata
SPICY 89018 & $1.35\pm0.18$ & $0.54\pm0.06$ & $0.21\pm0.05$ & $0.08\pm0.03$\\
SPICY 89954 & $1.51\pm0.16$ & $0.47\pm0.08$ & $0.23\pm0.05$ & $0.15\pm0.04$\\
SPICY 90918 & $1.97\pm0.25$ & $0.78\pm0.10$ & $0.13\pm0.06$ & $0.12\pm0.03$\\
SPICY 90923 & $0.32\pm0.12$ & $0.08\pm0.03$ & $0.08\pm0.06$ & $0.06\pm0.02$\\
SPICY 93027 & $2.50\pm0.32$ & $0.68\pm0.09$ & $0.41\pm0.09$ & $0.12\pm0.03$\\
SPICY 98558 & $1.59\pm0.16$ & $0.48\pm0.08$ & $0.18\pm0.06$ & $0.17\pm0.04$\\
SPICY 100086 & $0.86\pm0.17$ & $0.24\pm0.03$ & $0.07\pm0.03$ & $0.10\pm0.03$\\
SPICY 103533 & $1.32\pm0.11$ & $0.48\pm0.02$ & $0.14\pm0.03$ & $0.16\pm0.04$\\
SPICY 104101 & $2.28\pm0.18$ & $0.85\pm0.08$ & $0.33\pm0.07$ & $0.31\pm0.05$\\
SPICY 105586 & $0.85\pm0.12$ & $0.28\pm0.06$ & $0.17\pm0.04$ & $0.06\pm0.03$\\
SPICY 105733 & $1.24\pm0.11$ & $0.53\pm0.08$ & $0.19\pm0.08$ & $0.23\pm0.06$\\
SPICY 105879 & $0.66\pm0.18$ & $0.20\pm0.09$ & $0.22\pm0.07$ & $0.11\pm0.05$\\
SPICY 106482 & $1.37\pm0.16$ & $0.61\pm0.10$ & $0.21\pm0.06$ & $0.07\pm0.04$\\
SPICY 106859 & $\ldots$ & $0.20\pm0.06$ & $0.12\pm0.09$ & $\ldots$\\
SPICY 107233 & $2.50\pm0.22$ & $1.16\pm0.06$ & $0.27\pm0.07$ & $0.27\pm0.05$\\
SPICY 108375 & $\ldots$ & $0.46\pm0.09$ & $0.26\pm0.10$ & $0.16\pm0.09$\\
SPICY 108400 & $1.02\pm0.13$ & $0.41\pm0.07$ & $\ldots$ & $0.25\pm0.05$\\
SPICY 108560 & $\ldots$ & $0.19\pm0.03$ & $0.13\pm0.04$ & $0.09\pm0.03$\\
SPICY 111557 & $1.23\pm0.59$ & $0.84\pm0.14$ & $0.34\pm0.11$ & $0.22\pm0.08$\\
SPICY 111583 & $1.27\pm0.27$ & $1.00\pm0.06$ & $0.10\pm0.07$ & $0.37\pm0.06$\\
SPICY 113327 & $0.57\pm0.11$ & $0.10\pm0.08$ & $0.13\pm0.08$ & $\ldots$\\
SPICY 115897 & $\ldots$ & $0.53\pm0.16$ & $\ldots$ & $\ldots$\\
SPICY 116344 & $1.83\pm0.34$ & $0.28\pm0.10$ & $\ldots$ & $0.12\pm0.04$\\
SPICY 116390 & $\ldots$ & $0.13\pm0.07$ & $0.11\pm0.06$ & $\ldots$\\
SPICY 116475 & $2.12\pm0.17$ & $0.61\pm0.08$ & $0.19\pm0.04$ & $0.15\pm0.06$\\
SPICY 117231 & $1.12\pm0.17$ & $0.39\pm0.08$ & $0.11\pm0.04$ & $0.15\pm0.06$\\
\enddata
\tablecomments{DIB equivalent widths are measured in spectral ranges individually selected for each DIB and each star based on visual inspection of the line.}
\end{deluxetable}

\end{document}